\documentclass[a4paper]{amsart}
\usepackage{amscd}
\def\opn#1 {\operatorname{#1}}
\def\dopn#1 {
 \def\mYname{\operatorname{#1}}
 \expandafter\let\csname#1\endcsname=\mYname}

\def\br#1{
 \ifx#1<\gdef\Br##1>{\left<##1\right>}\else
 \ifx#1(\gdef\Br##1){\left(##1\right)}\else
 \ifx#1[\gdef\Br##1]{\left[##1\right]}\else
 \ifx#1\{\gdef\Br##1\}{\left\{##1\right\}}\else
 \ifx#1|\gdef\Br##1|{\left|##1\right|}\else
 \ifx#1\|\gdef\Br##1\|{\left\|##1\right\|}\else
 \errmessage{\string\br:****Bad bracket!****}\fi\fi\fi\fi\fi\fi
 \Br}

\let\too=\xrightarrow
\let\ltoo=\xleftarrow
\def\seq{\subseteq}
\def\cj{\overline}
\def\ul{\underline}

\newtheorem{thm}{Theorem}[subsection]
\newtheorem{cor}[thm]{Corollary}
\newtheorem{lem}[thm]{Lemma}
\newtheorem{prp}[thm]{Proposition}
\theoremstyle{remark}
\newtheorem{rmn}[thm]{Remark} 
\newtheorem{rmns}[thm]{Remarks} 
\newenvironment{rem}{\par\smallskip\par{\em Remark.}}{\par\smallskip\par}
\newenvironment{rems}{\par\smallskip\par{\em Remarks.}}{\par\smallskip\par}

\long\def\CAR#1#2\NIL{#1}
\long\def\Brm#1\Erm{
 \edef\nxt{\CAR#1\relax\NIL}
 \expandafter\ifx\nxt(
 \begin{rems} #1 \end{rems}\else
 \begin{rem} #1 \end{rem}\fi
}
\long\def\Brmn#1 #2\Erm{
 \edef\nxt{\CAR#2\relax\NIL}
 \expandafter\ifx\nxt(
 \begin{rmns}\label{#1} #2 \end{rmns}\else
 \begin{rmn}\label{#1} #2 \end{rmn}\fi
}

\numberwithin{equation}{subsection}
\def\nt{\\}
\def\Beq#1\Eeq{\begin{equation*} #1 \end{equation*}}
\def\Beqn#1 #2\Eeq{\begin{equation}#2 \label{#1} \end{equation}}
\def\Bml#1\Eml{\begin{multline*} #1 \end{multline*}}
\def\Bmln#1 #2\Eml{\begin{multline}#2 \label{#1} \end{multline}}
\def\Bal#1\Eal{\begin{align*} #1 \end{align*}}
\def\Baln#1 #2\Eal{\begin{align}\label{#1} #2 \end{align}}
\def\Bgt#1\Egt{\begin{gather*} #1 \end{gather*}}
\def\Bgtn#1 #2\Egt{\begin{gather}\label{#1} #2 \end{gather}}
\def\Bea#1\Eea{\begin{eqnarray*} #1 \end{eqnarray*}}
\def\Bean#1 #2\Eea{\begin{eqnarray} #2 \label{#1}\end{eqnarray}}
\def\Bcd#1\Ecd{\[\begin{CD} #1 \end{CD}\]}
\def\Bcdn#1 #2\Ecd{
 \begin{equation}\begin{CD}#2 \label{#1}\end{CD}\end{equation}}
\def\bysame{\leavevmode\hbox to3em{\hrulefill}\,} 

\DeclareMathSymbol{\Subset} {\mathrel}{AMSa}{"62}
\begin{document}
\dopn A
\def\even{{\mathbf0}}
\def\odd{{\mathbf1}}
\def\seven{_\even}
\def\sodd{_\odd}
\def\sevR{_{\even,\Bbb R}}
\def\sodR{_{\odd,\Bbb R}}
\def\O{\mathop{\mathcal O}\nolimits}
\def\M{\mathop{\mathcal M}\nolimits}
\def\P#1;{{\mathcal P}#1;\ }            
\def\Pf#1;{{\mathcal P}_{\opn f }#1;\ } 
\dopn L
\dopn E
\dopn CS
\dopn CB
\dopn e
\dopn i
\dopn pr
\dopn supp
\def\Cau{^{\opn Cau }}
\def\XiCauP{{\Xi'}^{\opn Cau }}
\def\sol{^{\opn sol }}
\def\free{^{\opn free }}
\def\cfg{^{\opn cfg }}
\def\zero{^{\opn zero }}
\def\ztwo{\Bbb Z_2}
\def\ext{^{\opn ext }}
\def\src{^{\opn src }}
\def\solsrc{^{\opn sol,J }}
\def\Xiexbar{{\Xi'}\ext}
\def\Xisrcbar{{\Xi'}\src}
\def\XiCData{\Xi\sol_\phi}
\newdimen\mYd  \newbox\mYbox
\let\dcj=\cj

\def\spsc{S} 
\def\mutarg{{\scriptstyle{\bullet}}}

\def\whatref#1{\cite[#1]{[WHAT]}}
\def\CMPref#1{\cite[#1]{[CMP1]}}
\def\CMPDefinesFGerm{\cite[Prop. 3.5.2]{[CMP1]}}
\def\CMPTransl{\cite[3.3]{[CMP1]}}
\def\CMPDefAdm{\cite[3.1]{[CMP1]}}
\def\sd{\tau} 
\def\smloss{\mu} 
\def\rdmm{\Bbb R^{d+1}}
\def\rdm{\Bbb R^d}
\def\trp{^{\opn T }} 
\def\cH{{\mathcal H}}
\def\cD{{\mathcal D}}
\def\V{^{\opn V }}
\def\cEV{{\mathcal E\V}}
\def\cECauV{{\mathcal E}^{\opn Cau,V }}
\def\cECauVs{(\cECauV)\seven}
\def\cEcV{{\mathcal E}\V_c}
\def\cEcCauV{{\mathcal E}_c^{\opn Cau,V }}
\def\cEcCauVs{(\cEcCauV)\seven}
\def\cEcCauVso{(\cEcCauV)\sodd}
\def\cEcVs{(\cEcV)\seven}
\def\Ci{{\mathcal C}\V_\infty}
\def\CiCau{{\mathcal C}_\infty^{\opn Cau,V }}
\def\ball#1{#1\Bbb B}
\def\causall{\prec}
\def\bV{{\Bbb V}}
\let\alb=\allowbreak
\def\J{{\mathcal J}}

\title[The Cauchy Problem]{
 The Cauchy Problem for Abstract Evolution Equations
\linebreak
 with Ghost and Fermion Degrees of Freedom
}
\author{T. Schmitt}
\begin{abstract}
We consider a class of abstract nonlinear evolution equations in supermanifolds
(smf's) modelled
over $\mathbb Z_2$-graded locally convex spaces. We show uniqueness, local
existence,
smoothness, and an abstract version of causal propagation of the solutions. If
an a-priori estimate
prevents the solutions from blowing-up then an infinite-dimensional smf of
"all" solutions
can be constructed.

We apply our results to a class of systems of nonlinear field equations with
anticommuting fields which arise in classical field
models used for realistic quantum field theoretic models.
In particular, we show that under suitable conditions,
the smf of smooth Cauchy data with compact support
is isomorphic with an smf of corresponding classical solutions of the model.
\end{abstract}
\maketitle

{
\def\linespacing{1pt}
\tableofcontents
}

\newpage

\section{Introduction and preliminaries}

\subsection{Introduction}
The investigation of the field equations belonging to a
quantum-field theoretical model as classical nonlinear wave equations
has a long history, dating back to Segal \cite{[SegalQW]}, \cite{[SegalNLSgr]};
cf. also
\cite{[ChoYM]}, \cite{[Eardley/Moncrief]}, \cite{[Ginibre/Velo]},
\cite{[Sniatycki]}.
Usually, Dirac fields have been considered in the obvious way as sections of a
spinor bundle, as e.~g. in \cite{[ChoYM]}.

On the other hand, the rise of supersymmetry made the question of an
adequate treatment of the fermion fields urgent --- supersymmetry and
supergravity do not work with commuting fermion fields. The same applies
to ghost fields: BRST symmetry, which now arouses a considerable interest
among mathematicians (cf. e. g. \cite{[KostBRST]}), simply does not exist with
commuting ghost fields.

The anticommutivity required from fermion and ghost fields is often
implemented by letting these fields have their values in the odd part of
an auxiliary Grassmann algebra, as e. g. in \cite{[ChoSugr]}.
However, in \cite{[WHAT]}, we have raised our objections against the use
of such an algebra, at least as a fundamental tool.

As we have argued in \cite{[WHAT]}, a satisfactory description
of fermion and ghost fields is possible in the framework of
infinite-dimensional supergeometry: the totality of configurations on
space-time should not be considered as a set but as an infinite-dimensional
supermanifold (smf), and the totality of classical solutions should be
a sub-supermanifold. While in \cite{[CMP1]}, \cite{[WHAT]}, we have developed
the necessary supergeometric machinery, this paper will combine it with old
and new techniques in non-linear wave equations in order to implement this
point of view.

Our motivating example is the standard Lagrangian of quantum chromodynamics,
which is a $\opn SU (3)$ Yang-Mills theory coupled with spinorial fields in the
fundamental
representation (in this paper, we will not really study any example;
a systematic application of our results to a large class of classical
field theories will be given in the successor paper):
\Beqn YMLagr
 {\mathcal L}[ A,\Psi] = -\frac14 
 F^{ab}F_{ab}
 + \frac\i2 
 \left(
  \dcj{\Psi}\gamma^a D^A_a\Psi - \dcj{D^A_a\Psi}\gamma^a\Psi\right)
  - m\dcj{\Psi}\Psi
\Eeq
with $F_{ab}[ A] :=\partial_b A_a-\partial_a A_b + \br[ A_a, A_b]$, and
$D^A_a:= \partial_a + \i/2  A_a^i\lambda_i$,\ \
$\lambda_1,\dots,\lambda_8$ are the Gell-Mann matrices which
realize the fundamental representation of $\opn su (3)$,
and we are using Einstein's summation convention, with  suppressing spinor
indices as well as the coupling constant. 

It is well-known that in order to get a well-posed Cauchy problem, we have
to break the gauge symmetry. Although this is rather unphysical, we pass
here to the temporal gauge $A_0=0$. (Unfortunately, this breaks Poincar\'e
invariance; for the treatment with gauge-breaking term and ghosts preferred
in physicist's textbooks, we do not yet have the necessary a priori
estimates to show completeness; cf. Thm. \ref{ComplThm}).

It is reasonable to conjecture that the arising equations of motion
\Beqn YMEqu
 \partial_a F^{ab}_i - \br[A_a,F^{ab}]_i =  \frac12
\dcj{\Psi}\gamma^b\lambda_i\Psi,
 \quad \i\gamma^aD^A_a\Psi= m\dcj{\Psi}
\Eeq
are all-time solvable. For the pure Yang-Mills case $\Psi=0$,
this is is already a highly non-trivial result proven in
\cite{[Eardley/Moncrief]}; cf. also
\cite{[Sniatycki]}.

Thus, for any $k>3/2$, let
\Beq
  B'_k := H_k(\Bbb R^3)\otimes\Bbb R^{24} \ \ \oplus \ \
             H_{k-1}(\Bbb R^3)\otimes\Bbb R^{24} \ \ \oplus \ \
            H_{k-1}(\Bbb R^3)\otimes\Bbb C^{12}
\Eeq
be the Banach space of Cauchy data $(A\Cau,\dot A\Cau,\Psi\Cau)$;
here $H_k$ is the usual Sobolev space $W^2_k$.
Also, let $C(\Bbb R,B'_k)$ denote the space of continuous functions $\Bbb R\to
B'_k$.
Then a precise formulation of the conjecture above states that
there should exist a map
\Beqn SolMap
 B'_k \to C(\Bbb R,B'_k),\qquad (A\Cau,\dot A\Cau,\Psi\Cau) \mapsto
(A\sol,\partial_tA\sol,\Psi\sol)
\Eeq
such that $(A\sol,\partial_tA\sol,\Psi\sol)|_{t=0}=(A\Cau,\dot
A\Cau,\Psi\Cau)$,
and the equations \eqref{YMEqu}
are satisfied. (It follows from the results presented below that this map, once
its existence can be proven,
will be uniquely determined, and in fact real-analytic, and thus the Cauchy
problem will have the best stability
property one can want.)

However, it is well-known (at least in the physical literature)
that the classical field $\Psi$ should be treated as {\em anticommuting},
i. e.
\Beqn AntiCom
  \br[\Psi_i(x),\Psi_j(y)]_+ =0
\Eeq
for all $x,y\in\Bbb R^4$ and indices $i,j$. This requirement is not
satisfied by modelling $\Psi$ as a function on space-time; in fact,
it drastically changes the meaning of \eqref{YMEqu}. It is even problematic
what a configuration should be.

As the author argued in \cite{[WHAT]}, the conceptually best answer to the
problem of satisfying \eqref{AntiCom} is the following:
the totality of configurations of the
classical fields should not be modelled as a set
(in our example the set $C(\Bbb R,B'_k)$)
but as an {\em infinite-dimensional supermanifold}. Roughly speaking, the
coordinates of this
supermanifold are the degrees of freedom of the model: the bosonic field
strengthes $A_a^n(x)$
for all $x\in\Bbb R^4$ are the even coordinates, the fermionic field strengthes
$\Psi_\alpha^n(x)$ are the odd ones. 

This implies that the meanwhile well-established framework of finitedimensional
supergeometry (cf. \cite{[Lei1]}, \cite{[Kostant]}, \cite{[HERM]}) has to be
extended to the infinite-dimensional case. (Cf. also \cite{[WHAT]} for a
discussion why we prefer the Berezin-Leites-Kostant approach to supermanifolds
to the deWitt-Rogers one.)

A calculus of real-analytic  supermanifolds (smf's) modelled over locally
convex spaces, suitable for our purposes, has been constructed
by the present author in \cite{[IS]}, \cite{[CMP1]}; cf. the remarks in
the next section. Thus, we replace $B'_k$ by the $\Bbb Z_2$-graded Banach space
\Beq
 B_k:=
 H_k(\Bbb R^3)\otimes\Bbb R^{24} \ \  \oplus \ \
 H_{k-1}(\Bbb R^3)\otimes\Bbb R^{24} \ \  \oplus \ \
 \Pi H_{k-1}(\Bbb R^3)\otimes\Bbb C^{12},
\Eeq
where, as usual in supergeometry, $\Pi$ is a formal odd symbol,
and we assign to it the corresponding {\em supermanifold of Cauchy data}
$\L(B_k)$, which is the linear (or "affine") supermanifold with model space
$B_k$. Also, the Fr\`echet space $C(\Bbb R,B_k)$ inherits a
$\Bbb Z_2$-grading, and the associated linear smf $\L(C(\Bbb R,B_k))$ is in our
approach the {\em supermanifold of configurations}.

Instead of the map \eqref{SolMap}, the results of this paper combined
with that of \cite{[Eardley/Moncrief]} yield a morphism
\Beqn APsiUnivSol
 (A\sol,\partial_t A\sol,\Psi\sol): \L(B_k)\to \L(C(\Bbb R,B_k))
\Eeq
such that \eqref{APsiUnivSol} solves \eqref{YMEqu}, and its time zero Cauchy
datum,
$(A\sol,\partial_t A\sol,\Psi\sol)(0)\in\O^{B_k}(\L(B_k))$, is just the
standard coordinate superfunction $(A\Cau,\dot A\Cau,\Psi\Cau)$. Moreover, it
turns out
that the image of \eqref{APsiUnivSol} exists as a sub-smf
$\L(C(\Bbb R,B_k))\sol\seq\L(C(\Bbb R,B_k))$; we call
$\L(C(\Bbb R,B_k))\sol$ the {\em supermanifold of classical solutions
of \eqref{YMEqu} within $\L(C(\Bbb R,B_k))$}.

However, viewing $\L(C(\Bbb R,B_k))\sol$ as "the" manifold of classical
solutions has the severe defect that we do not know whether it is Lorentz
invariant in a reasonable sense; probably, it is not. At any rate, there is no
reasonable action of the Lorentz group on $\L(C(\Bbb R,B_k))$. (Of course,
in this particular example, Lorentz invariance is spoiled anyway by the
temporal
gauge condition. But the objection stands for many other models.)

An obvious proposal for improvement is to use smooth Cauchy data and
configurations. Thm. \ref{MainThmSm} below yields the following variant of
\eqref{APsiUnivSol}:
\Bml
 (A\sol,\partial_t A\sol,\Psi\sol):
\L\bigl(C^\infty(\Bbb R^3)\otimes\Bbb R^{48}
 \  \oplus \ \Pi C^\infty(\Bbb R^3)\otimes\Bbb C^{12}\bigr) \to
\\
\to\L\bigl(C^\infty(\Bbb R^4)\otimes\Bbb R^{24}
 \  \oplus \ \Pi C^\infty(\Bbb R^4)\otimes\Bbb C^{12}\bigr).
\Eml
(Actually, in order to derive this, one has to use the formulation of the
Yang-Mills equation
given originally by Segal, since that used by \cite{[Eardley/Moncrief]},
although
better reflecting the degrees of smoothness, obscures the causal properties. A
systematic discussion
will be given in a successor paper.)

Again, this possesses an image sub-smf, the {\em smf of
smooth solutions of \eqref{YMEqu}}.

However, while the absence of any growth condition in spatial direction does
not cause trouble in the construction, due to finite propagation speed,
it causes difficulties in the subsequent investigation of
differential-geometric structures on the image $M\sol_{C^\infty}$:
Roughly spoken, any superfunction $K[\Phi|\Psi]$ on the Cauchy smf
is influenced only by the "values" of the fields
on the finite region $\Omega$. In particular, the energy at a given
time instant is not a well-defined superfunction;
only the energy in a finite space-time region is so.
What is still worse, the symplectic structure on
the solution smf which one expects (cf. \whatref{1.12.4}),  and which we will
study in subsequent papers, simply does not make sense; only the
corresponding Poisson structure does.

Thus, it seems reasonable to use only smooth Cauchy data with compact support,
i. e. of test function quality. However, we have to be careful in the
choice of the model space for the target smf: simply taking all smooth
functions on $\Bbb R^4$ which are spatially compactly supported would violate
Lorentz invariance. However, if we additionally suppose that the spatial
support grows only with light speed then everything is OK:
Let $C^\infty_c(\Bbb R^4)$
denote the space of all $f\in C^\infty(\Bbb R^4)$
such that there exists $R>0$ with $f(t,x) = 0$ for all
$(t,x)\in\Bbb R\times\Bbb R^3$ with $\br|x|\ge \br|t| + R$.
(Note that this is only a strict inductive limes
of Fr\`echet spaces.) Thm. \ref{cDMainThm} now yields that
\eqref{APsiUnivSol} restricts to a morphism
\Bml
 (A\sol,\partial_t A\sol,\Psi\sol):
\L\bigl(C^\infty_0(\Bbb R^3)\otimes\Bbb R^{48}
 \  \oplus \ \Pi C^\infty_0(\Bbb R^3)\otimes\Bbb R^{12}\bigr) \to
\\
\to
\L\bigl(C^\infty_c(\Bbb R^4)\otimes\Bbb R^{24}
 \  \oplus \ \Pi C^\infty_c(\Bbb R^4)\otimes\Bbb R^{12}\bigr)
\Eml
Again, this possesses an image sub-smf, the {\em smf of
smooth solutions of \eqref{YMEqu} with causally growing spatially compact
support}. In a subsequent paper, we will show that for suitable models,
this smf is acted upon by the Poincar\'e group and carries an invariant
symplectic structure.

\subsection{Infinite-dimensional supergeometry}\label{InfDimSGeom}
Let us shortly recall some notions and conventions from
\cite{[CMP1]}, \cite{[WHAT]}. We follow the usual conventions of
$\Bbb Z_2$-graded algebra: All vector spaces will be $\Bbb Z_2$-graded,
$E=E\seven\oplus E\sodd$ (decomposition into {\em even} and {\em odd} part);
for the {\em parity} of an element, we will write $\br|e|=\mathbf i$ for
$e\in E_{\mathbf i}$. In multilinear expressions, parities add up; this fixes
parities for tensor product and linear maps. (Note that space-time, being not
treated as vector space, remains ungraded. On the other hand, "classical"
function spaces, like Sobolev spaces, are treated as purely even.)

{\em First Sign Rule:} Whenever in a complex multilinear expression
two adjacent terms $a,\ b$ are interchanged the sign $(-1)^{\br|a|\br|b|}$
has to be introduced.

In order to get on the classical level a correct model of operator
conjugation in the quantized theory we also have to use the additional
rules of the hermitian calculus developed in \cite{[HERM]}.
That is, the role of real supercommutative algebras is taken over by
{\em hermitian supercommutative algebras}, i.~e.
complex supercommutative algebras $R$ together with
an involutive antilinear map $\cj{\cdot}: R\to R$
({\em hermitian conjugation\/}) such that $\cj{rs}=\cj s\cdot\cj r$
for $r,s\in R$ holds. Note that this rule does not contradict the first sign
rule since
$\cj{rs}$ is not complex multilinear in $r,s$. Also,
{\em the real elements of a hermitian
algebra do in general not form a subalgebra}, i. e. $R$ is not just the
complexification of a real algebra. More general, all real vector spaces
have to be complexified before its elements may enter multilinear
expressions. The essential ingredient of the hermitian framework is the

{\em Second Sign Rule:} If conjugation is applied to a bilinear
expression in the terms $a,\ b$ (i.~e. if conjugation is resolved
into termwise conjugation), either $a,\ b$ have to be rearranged
backwards, or the expression acquires the sign factor $(-1)^{\br|a|\br|b|}$.
Multilinear terms have to be treated iteratively.

A calculus of real-analytic infinite-dimensional
supermanifolds (smf's) has been constructed by the present author in
\cite{[IS]}, \cite{[CMP1]}.
Here we note that it assigns to every real $\Bbb Z_2$-graded locally
convex space (henceforth abbreviated $\Bbb Z_2$-lcs) $E=E\seven\oplus E\sodd$ a
{\em linear
supermanifold} $\L(E)$ which is essentially
a ringed space $\L(E)=(E\seven,\O)$ with underlying topological space
$E\seven$ while the structure
sheaf $\O$ might be thought very roughly of as a kind of completion of
${\mathcal A}(\cdot)\otimes \Lambda E^*_{\odd,\mathbb C}$; here
${\mathcal A}(\cdot)$ is the sheaf of real-analytic functions on the even part
$E\seven$
while $\Lambda E^*_{\odd,\mathbb C}$ is the exterior
algebra over the complexified dual of the odd part of $E$.

The actual definition of the structure sheaf treats even and odd sector
much more on equal footing than the tensor product ansatz above:
Given a second real $\Bbb Z_2$-lcs $F$, one defines the $\Bbb Z_2$-graded
complex vector space
$\Pf(E;F)$ of {\em $F$-valued formal power series on $E$} as the set of all
formal
sums $u=\sum_{k,l\ge0} u_{(k|l)}$ where
$u_{(k|l)}: \prod^k E\seven \times \prod^l E\sodd \to F\otimes_{\Bbb R} \Bbb C$
is a jointly continuous, multilinear map which is symmetric
on $E\seven$ and alternating on $E\sodd$.  This space has a natural hermitian
conjugation, and,
by usual multilinear techniques, one constructs an associative bilinear pairing
$\Pf(E;F) \times \Pf(E;F') \to \Pf(E;F\otimes F')$; in particular, $\Pf(E;\Bbb
R)$ becomes a
$\Bbb Z_2$-commutative hermitian algebra.

Recall that, assigning to a seminorm
$p$ its unit ball $\{e\in E:\ p(e)\le1\}$, we get a bijection between
the set $\CS(E)\owns p$ of continuous seminorms on $E$, and the
set $\CB(E)$ of convex balanced closed neighbourhoods of the origin.

Now let $F$ be a $\Bbb Z_2$-graded Banach space, and $U\in\CB (E)$.
For $u\in\P(E;F)$, let $\br\|u_{(k|l)}\|$ be the supremum of
$\br\|u_{(k|l)}(\cdot)\|$ on
$\prod^k (U\cap E\seven) \times \prod^l  (U\cap E\sodd)$.
Let $\P(E,U;F)$ be the Banach space of all those  $u\in\Pf(E;F)$ for which
$\br\|u\|:= \sum_{k,l\ge0}\br\|u_{(k|l)}\|$ is finite.
Conforming with \cite{[CMP1]}, we will denote this space also by $\P(E,p;F)$
where $p$ is the
seminorm with unit ball $U$.

Conceptually, $\P(E,U;F)$ is the {\em space of power series converging on $U$}.
Indeed,
every element $K\in\P(E,U;F)$ is "a function element on $U\cap E\seven$", i.~e.
it will be
the Taylor expansion at zero of a uniquely determined superfunction
$K\in\O^F(U\cap E\seven)$ within the superdomain $\L(E)$ (cf.
\CMPDefinesFGerm).

Define the {\em space $\P(E;F)$ of analytic power series from $E$ to $F$} as
the set of
all $u\in\Pf(E;F)$ such that for all $p\in\CS(F)$ there exists $U\in\CB(E)$
such that
$i_p\circ u\in \P(E,U;\hat F_p)$ where $i_p:F\to\hat F_p$ is the canonical map
into
the completion of $F$ w.~r. to $p$ (with the zero space of $p$ factored out).

Given power series $u\in\P(E,U;F)$ where $F$ is Banach and
$v\in\P(E';E)_{\even,\Bbb R}$
with $v_{(0|0)}\in U$, one defines with some multilinear voodoo the {\em
composition}
$u[v]\in\P(E';F)$; cf. \whatref{2.3} for details.

Now, for any  $\Bbb Z_2$-lcs $F$, one defines the
sheaf $\O^F(\cdot)$ of {\em $F$-valued superfunctions} on $E\seven$: an
element of $\O^F(U)$ where $U\seq E\seven$ is open is a map $f:U\to\P(E;F)$,
$e\mapsto f_e$, which satisfies a certain "coherence" condition which makes it
sensible
to interpret $f_e$ as the Taylor expansion of $f$ at $e$: One requires that
for all $p\in\CS(F)$ there exists $U\in\CB(E)$ such that
$i_p\circ f_{e+e'}[x] =i_p\circ f_e[x+e]$ for $e\in U\cap E\seven$. Here
$x\in\P(E;E)$ is the identity $E\to E$ viewed as power series; it acts as
identity under
composition.

Now the structure sheaf of our ringed space $\L(E)$ is simply
$\O(\cdot):=\O^{\Bbb R}(\cdot)$; it is a sheaf of hermitian supercommutative
algebras,
and each $\O^F(\cdot)$ is a module sheaf over $\O(\cdot)$.

Actually, in considering more general smf's than superdomains,
one has to enhance the structure of a ringed space slightly, in
order to avoid "fake  morphisms". What matters here is that the
enhancement is done in such a way that the following holds
(cf. \whatref{Thm. 2.8.1}):

\begin{lem} \label{CoordLem}
Given an $\Bbb Z_2$-lcs $F$ and an arbitrary smf $Z$, the set of morphisms
$Z\to\L(F)$ is in natural 1-1-correspondence with the set
\Beq
 \M^F(Z):= \O^F(Z)_{\even,\Bbb R}.
\Eeq
(Here the subscript stands for the real, even part.)
The correspondence works as follows: There exists a distinguished element
$x\in\M^F(\L(F))$ called the {\em standard coordinate}, and one assigns to
$\mu:Z\to\L(F)$ the pullback $\mu^*(x)$.
\qed\end{lem}

(In previous papers, we had denoted this pullback by $\hat\mu$; in this one,
we will abuse notation and drop the hat, thus identifying a superfunction
$\mu\in\M^F(Z)$ with its corresponding morphism $\mu:Z\to\L(F)$.)

This is the infinite-dimensional version of the fact that if
$F=\Bbb R^{m|n}= \Bbb R^m\oplus\Pi\Bbb R^n$ is the standard $m|n$-dimensional
super vector space
then a morphism $Z\to\L(\Bbb R^{m|n})$ is known by knowing
the pullbacks of the coordinate superfunctions, and these can be
prescribed arbitrarily as long as parity and reality are OK (cf. e. g.
\cite[Thm. 2.1.7]{[Lei1]}).

The most straightforward way to do the enhancement mentioned is a chart
approach; since the supermanifolds we are going to use are actually
all superdomains, and only the morphisms between them are non-trivial, we
need not care here for details.

If $E$, $F$ are spaces of generalized functions on $\rdm$ which contain
the test functions as dense subspace then the Schwartz kernel theorem tells
us that the multilinear forms $u_{(k|l)}$ are given by their integral
kernels, which are generalized functions. Thus one can apply
rather suggestive integral writings (cf. \cite{[CMP1]}) like e.~g.
\eqref{YMLagr}: The general form of a power series becomes
\Bmln GenPowSer
 K[\Phi|\Psi] = \sum_{k,l\ge0} \frac1{k!l!} \sum_{I,J}
 \int_{\Bbb R^{d(k+l)}} dx_1\cdots dx_kdy_1\cdots dy_l\cdot
 \\
 \cdot K^{i_1,\dots,i_k|j_1,\dots,j_l}(x_1,\dots,x_k|y_1,\dots,y_l)
 \Phi_{i_1}(x_1)\cdots\Phi_{i_k}(x_k)\Psi_{j_1}(y_1)\cdots\Psi_{j_l}(y_l)
\Eml
where we have used collective indices $i=1,\dots,N\seven$ and
$j=1,\dots,N\sodd$ for the real components of bosonic and fermionic
fields, respectively. The {\em coefficient functions}
$K^{i_1,\dots,i_k|j_1,\dots,j_l}(x_1,\dots,x_k|y_1,\dots,y_l)$
are distributions which can be supposed to be
symmetric in the pairs $(x_1,i_1),\dots,(x_k,i_k)$ and antisymmetric in
$(y_1,j_1),\dots,(y_l,j_l)$. Of course, they have to satisfy also certain
growth and smoothness conditions. However, what matters here is that the
$\Phi$'s and $\Psi$'s can be formally treated as commuting and anticommuting
fields, respectively; in fact, after establishing the proper calculational
framework, the writing \eqref{GenPowSer} is sufficiently correct.
Also, it is possible to substitute power series into each other
under suitable conditions. Cf. \cite{[CMP1]} for a detailed exposition.

We conclude with some additional preliminaries.
It will be convenient to work not with the bidegrees $(k|l)$
of forms but with {\em total degrees}: For any formal power series
$K\in\P_f(E;F)$ set for $m\ge0$
\Beq
 K_{(m)} :=\sum_{k=0}^m K_{(k|m-k)},\qquad
 K_{(\le m)} :=\sum_{n=0}^m K_{(n)}.
\Eeq
Thus $K=\sum_{m\ge0}K_{(m)}$.

Let $B$ be a $\Bbb Z_2$-graded Banach space and $E$ any $\Bbb Z_2$-lcs.
We call a superfunction $f\in\O^E(\L(B))$ {\em entire} if for every
$q\in\CS(E)$ and every $n>0$ we have $f_0\in\P(B,nU;\hat E_q)$ where
$f_0$ is the Taylor expansion at zero, and $U$ is the unit ball.
For instance, every $k|l$-form $u_{(k|l)}\in \Pf(B;E)$
is the Taylor expansion at zero of a unique entire superfunction.

\section{Results in the abstract setting}\label{ResAbstrSet}
\subsection{Configuration families}\label{ConfFam}

Through the whole section \ref{ResAbstrSet}, we fix a real $\Bbb Z_2$-graded
Banach space $B$ and a strongly continuous
group $(\A_t)_{t\in\Bbb R}$ of parity preserving bounded linear operators;
let $K: \opn dom K\to B$ denote the generator of this group.
Also, let be given an entire even, real superfunction
$\Delta\in \M^B(\L(B))$ the Taylor expansion of which in zero has lower degree
$\ge2$. Formally, the equation of interest is
\Beqn AbstrDEq
 \frac d{dt} \Xi'=K\Xi' + \Delta[\Xi'];
\Eeq
however, this makes sense only if $\Xi'$ takes values in $\opn dom K$.
Therefore we look for the integrated version
\Beqn AbstrIEq
 \Xi'(t) = \A_t\Xi'(0) + \int_0^t ds \A_{t-s}\Delta[\Xi'(s)].
\Eeq
Before embarking in \ref{SolFam} into the explanation of the precise meaning of
this equation, we first have to clarify the meaning of $\Xi'$.

For a connected subset $I\seq \Bbb R$, $I\owns0$,  with non-empty open kernel,
let $B(I):=C(I,B)$ equipped with the topology induced by the seminorms
$\br\|\xi\|_{B([a,b])}:= \max_{t\in[a,b]} \br\|\xi(t)\|$ where
$a,b\in I$, $a<b$.

Let $Z$ be an arbitrary smf. A {\em configuration family
parametrized by $Z$ with time definition domain $I$}
(or {\em $Z$-family}, for short) is an even, real
superfunction $\Xi'$ on $Z$ with values in the locally convex space
$B(I)$:
\Beq
 \Xi'\in\M^{B(I)}(Z)
\Eeq
(we recall that $\M$ denotes the real, even part of the sheaf $\O$).
Now, given an smf morphism $\pi:Z'\to Z$ we can assign to every
$Z$-family $\Xi'$ its {\em pullback} $\Xi":=\pi^*(\Xi')$ which
is a $Z'$-family. In fact, the process of passing from $\Xi'$ to
$\Xi"$ means in family language nothing but a {\em change of parametrization}
(cf. \whatref{1.11}).

\Brm
If $Z$ has odd dimension zero, i.~e. is effectively an ordinary manifold,
then $\Xi'$ is the same as a real-analytic map $\Xi':Z\to B\seven$.
In particular, this applies if $Z=P$ is a point; then a $Z$-family of solutions
is just
an element $\phi\in B(I)\seven$. If also $B=B\seven$ is purely even, so that
the problem \eqref{AbstrIEq} is a classical one, then this is a configuration
in the usual sense.

However, if $B\sodd\not=0$, it follows that configuration families with
non-trivial
odd sector are necessarily parametrized by supermanifolds with non-vanishing
odd dimension; in particular, there are no "individual" configurations besides
purely
even ones.
\Erm

By Lemma \ref{CoordLem}, the smf $\L(B(I))$ is the moduli space for
configuration families
with time definition domain $I$, with the standard coordinate
\Beq
 \Xi\in\M^{B(I)}(\L(B(I)))
\Eeq
being the {\em universal configuration family with time definition domain $I$}.
Indeed, given an arbitrary $Z$-family $\Xi'\in\M^{B(I)}(Z)$ ,
it defines by Lemma \ref{CoordLem} a {\em classifying morphism}
$\Xi': Z\to \L(B(I))$, and $\Xi'$ arises from $\Xi$ just by pullback:
$\Xi'={\Xi'}^*(\Xi)$.

The {\em family of Cauchy data} of a family $\Xi'\in\M^{B(I)}(Z)$
with $I\owns0$ is the element $\Xi'(0)\in\M^B(Z)$.

\Brm In the language of category theory, this means that the
cofunctor
\Beq
 \{\text{supermanifolds}\}\to\{\text{sets}\},\qquad Z\mapsto\M^{B(I)}(Z),
\Eeq
is represented by the object $\L(B(I))$ with the universal element $\Xi$.
\Erm

\subsection{Solution families}\label{SolFam}

We turn to the r.~h.~s. of \eqref{AbstrIEq}: Let $\Xi'\in\M^{B(I)}(Z)$ be a
$Z$-family.
Although $\Delta[\Xi'(s)]\in \M^B(Z)$ is well-defined for each $s\in I$, we
still have to show well-definedness of the integral. We note that, as a
consequence
of Lemma \ref{DeltaLem} below, there exists a unique superfunction
$\Delta \in \M^{B(I)}(\L(B(I)))$ (by abuse of notation)
such that $\Delta[\Xi'](t) = \Delta[\Xi'(t)]$ for all $t\in I$.
Therefore $\Delta[\Xi']\in \M^{B(I)}(Z)$ is well-defined, and the
integral is now simply applied in the target.

It follows that the superfunction symbolically denoted by
\Beq
 \A_\mutarg\Xi'(0) + \int_0^\mutarg ds \A_{\mutarg-s}\Delta[\Xi'(s)]
 \in\M^{B(I)}(Z)
\Eeq
is well-defined (here and in the following, the small bullet stands for
the mute time argument).

We call $\Xi'$ a {\em $Z$-family of solutions},
or {\em solution family} for short, if \eqref{AbstrIEq} is satisfied.
Trivially, every pullback of a solution family is a solution family.

Thus, to solve the long-time Cauchy problem for the equation
\eqref{AbstrIEq} with a given family of Cauchy data $\XiCauP\in\M^B(Z)$
means to find a solution
family $\Xi'\in\M^{B(\Bbb R)}(Z)$ with $\Xi'(0)= \XiCauP$.

It turns out that it is sufficient to solve this problem for just one
universal family of Cauchy data, in oder to deduce solutions for all
other families of Cauchy data:

We call $\L(B)$ the {\em smf of Cauchy data}, and we denote its standard
coordinate by $\Xi\Cau\in \M^B(\L(B))$ and call it the {\em universal family
of Cauchy data}. Now suppose we have constructed a solution family
$\Xi\sol\in\M^{B(\Bbb R)}(\L(B))$ with $\Xi\sol(0)=\Xi\Cau$. Then, given an
other family of Cauchy data $\XiCauP\in\M^B(Z)$, it determines
by Lemma \ref{CoordLem} a morphism $Z \too{\XiCauP} \L(B)$, i.~e. it is
the pullback of the universal family of Cauchy data along this morphism.
The solution family $\Xi'$ sought for is then nothing but the pullback
of $\Xi\sol$ along this same morphism: $\Xi' = (\XiCauP)^*(\Xi\sol)$.

This justifies it to call $\Xi\sol$ the {\em universal solution family}, and
its construction is the main concern of this paper.

In this way, we arrive at the picture sketched in the Introduction.

If $Z=P$ is a point then a $Z$-family of solutions is just an element
$\phi\in B(I)\seven$ which solves the {\em underlying even problem}
\Beqn UnderlEvProbl
 \phi(t) = \A_t|_{B\seven}\phi(0) +
 \int_0^t  ds \A_{t-s}|_{B\seven}\tilde\Delta[\phi(s)]
\Eeq
in the usual sense.
($\tilde\Delta$ is the underlying function of the superfunction $\Delta$).

\begin{rem}
If $\Xi'$ takes values in $D:=\opn dom K$ (i.~e.
$\Xi'\in\M^{C(I,\opn dom K)}(Z)$ where $\opn dom K$ is equipped with
the graph norm) then \eqref{AbstrIEq} is equivalent to the differentiated
form \eqref{AbstrDEq}).
\end{rem}

Our first non-trivial result is Cauchy uniqueness
(in fact, this will be a special case of the more general result
Thm. \ref{CausUniqThm}):

\begin{thm}\label{UniqCor}
Let be given solution families $\Xi',\Xi"\in\M^{B(I)}(Z)$
such that for some $t_0\in I$ we have $\Xi'(t_0)=\Xi"(t_0)$.
Then $\Xi'=\Xi"$.
\end{thm}

All proofs, as far as not omitted, will be given in section
\ref{Proofs}.


\subsection{Completeness and solvability}

Loosely said, we call the problem \eqref{AbstrIEq} complete iff the
underlying even problem is all-time solvable; it is a standard
observation in nonlinear evolution equations that this is equivalent
with the existence of a-priori estimates. It turns out that this condition
also completely controls the solvability of the all-time Cauchy problem
for solution families:

\begin{thm}\label{ComplThm}
Fix the problem \eqref{AbstrIEq} and a
subset $A\seq B\seven$. The following conditions are equivalent:

(i) For every solution $\phi\in B((a,b))\seven$ of the underlying even
problem \eqref{UnderlEvProbl} on a bounded open time interval $(a,b)
\owns 0$ such that $\phi(0)\in A$, we have
\Beq
\sup_{t\in(a,b)} \br\|\phi(t)\|_B < \infty.
\Eeq

(ii) The underlying even problem is all-time solvable for Cauchy data
in $A$:
\begin{quotation}
Given Cauchy data $\phi\Cau \in A$ there exists an element $\phi\in
B(\Bbb R)$ with these Cauchy data which solves the problem
\eqref{UnderlEvProbl}.
\end{quotation}

(iii) The problem \eqref{AbstrIEq} is all-time solvable for
families of Cauchy data "taking values in $A$":

\begin{quotation}
Whenever we are given an smf $Z$ and an smf morphism
$\XiCauP:Z\to\L(B)$ such that
the image of the underlying map $\widetilde{\XiCauP}:\opn space (Z)\to
B\seven$
is contained in $A$, there exists a (necessarily uniquely determined)
solution family $\Xi'\in\M^{B(\Bbb R)}(Z)$ with $\Xi'(0) = \XiCauP$.
\end{quotation}

\noindent If these conditions are satisfied we call the the problem
\eqref{AbstrIEq} {\em $(B,A)$-complete}.
If the problem \eqref{AbstrIEq}
is $(B,B\seven)$-complete we call it simply {\em $B$-complete}.
\end{thm}

\Brm
(1) The notion "completeness" has been chosen by analogy with the
usual completeness of flows (i. e. local one-parameter groups of
automorphisms) on manifolds. Indeed, the problem \eqref{AbstrIEq}
determines a time evolution flow on the smf $\L(B)$, and it is
$B$-complete iff this flow is complete.

However, the problem in making that rigorous is that our smf calculus
is real-analytic while this flow in time direction is not.

(2) Let us comment on the fact that completeness depends only on the
underlying even problem \eqref{UnderlEvProbl}: Mathematically, this is an
analogon of several theorems in supergeometry that differential-geometric
tasks, like trivializing a fibre bundle, or presenting a closed form
as differential, are solvable iff the underlying smooth tasks are solvable.

Physically, our interpretation is somewhat speculative:
In the bosonic sector, the classical field theory approximates the behaviour
of coherent states, and completeness excludes that "too many" particles may
eventually assemble at a space-time point, making the state non-normable.
On the fermionic side, apart from the non-existence of genuine coherent
states, it is the Pauli principle which automatically prevents such an
assembly.
\Erm

\subsection{Solvability in function spaces}

We will be interested in subspaces of $B(\Bbb R)$ which are determined by
additional functional-analytic quality conditions. Thus, suppose that
we are given a continuous, even inclusion $E\seq B(\Bbb R)$ where
$E$ is another $\ztwo$-lcs.

We call any element $\Xi'\in\M^{E}(Z)$ a {\em $Z$-family of configurations
of quality $E$}.
By the inclusion $\M^{E}(Z)\seq \M^{B(\Bbb R)}(Z)$, such an element can be
viewed as a $Z$-family in the previous sense, and so it makes sense to
speak of {\em solution families of quality $E$}.

Denote by $E\Cau\seq B$ the image of $E$ under the projection onto Cauchy data,
and equip it with the quotient topology. We call $E\Cau$ the
{\em space of Cauchy data belonging to $E$}. 

\label{DefSolvable}
We call the problem \eqref{AbstrIEq} {\em solvable in $\L(E)$} if:
there exists a (necessarily uniquely determined) solution family
of quality $E$ henceforth denoted by $\Xi\sol\in\M^E(\L(E\Cau))$ the
Cauchy data of which is the standard coordinate
$\Xi\Cau\in \M^{E\Cau}(\L(E\Cau))$:
\Beq
 \Xi\sol(0) = \Xi\Cau.
\Eeq

\begin{thm}\label{IsSubSmf}
If the problem \eqref{AbstrIEq} is solvable in $\L(E)$
then the image of the morphism $\Xi\sol: \L(E\Cau)\to\L(E)$ is a
split sub-smf which we call the {\em smf of classical solutions
within $\L(E)$}, and denote by $\L(E)\sol\seq \L(E)$.
\end{thm}

$\L(E)\sol$ has the following universal property:
Fixing an smf $Z$ we have a bijection between $Z$-families
$\Xi'$ of configurations of quality $E$ with time definition
interval $\Bbb R$, and morphisms
$\Xi':Z\to \L(E)$. Now $\Xi'$ is a solution
family iff the morphism $\Xi'$ factors to
$\Xi':Z\to \L(E)\sol\seq \L(E)$.

In this way, we get a bijection between $Z$-families
$\Xi'$ of solutions of quality $E$ with time definition interval
$\Bbb R$, and morphisms $\Xi':Z\to \L(E)\sol$.

\Brm
(1) Suppose that the problem \eqref{AbstrIEq} is both solvable in $\L(E)$
and in $\L(B(\Bbb R))$. Then we have a commutative diagram
\Bcd
 \L(E\Cau) @>\Xi\sol>> \L(E)\nt
 @V VV        @V VV \nt
 \L(B) @>\Xi\sol>> \L(B(\Bbb R)),
\Ecd
which justifies it to use the same notation $\Xi\sol$ in all cases.
Moreover, $\L(E)\sol$ is just the intersection $\L(E)\cap \L(B(\Bbb R))\sol$
in the categorial sense, i.~e. the pullback of the diagram
$\L(E) \too\seq \L(B(\Bbb R)) \ltoo\supseteq \L(B(\Bbb R))\sol$.

(2) It follows that the underlying manifold $\widetilde{\L(E)\sol}$
identifies with the set of all $\phi\in E\seven$ which satisfy
\eqref{UnderlEvProbl}.

(3) Note that $\L(E)\sol$ is still a linear smf which is, however, in
a non-linear way embedded into $\L(E)$.
\Erm

An obvious necessary condition for solvability in $\L(E)$ is
$(B,(E\Cau)\seven)$-completeness.
For the maximal choice $E=B(\Bbb R)$, it follows from Thm. \ref{ComplThm}
that this condition is also sufficient:

\begin{cor}\label{CorSolvInLBR}
The problem \eqref{AbstrIEq} is $B$-complete iff it is solvable in
$\L(B(\Bbb R))$.
\qed\end{cor}

Cor. \ref{GenSolvCrit} below gives a general method for showing solvability.

A simple but useful observation is:

\begin{cor}\label{ProjLim}
Let be given a family $(E_\kappa)_{\kappa\in K}$ of $\ztwo$-lcs
and continuous, even inclusions $E_\kappa\seq B(\Bbb R)$ such that
the problem \eqref{AbstrIEq} is solvable in each $\L(E_\kappa)$. Let
$E:=\bigcap_{\kappa\in K} E_\kappa$, equipped with the projective limes
topology. Then the problem \eqref{AbstrIEq} is solvable in $\L(E)$.
\qed\end{cor}

In the following, we will consider some special cases.


\subsection{Smoothness scales}

Up to now, the solutions the existence of which is asserted in Thm.
\ref{ComplThm}
and Cor. \ref{CorSolvInLBR} are in time direction only continuous. Using
smoothness
scales we get temporal differentiability properties.

Suppose we are given a sequence of real $\ztwo$-graded
Banach spaces and continuous even inclusions
\Beqn SmSc
 B=B_0\supseteq B_1 \supseteq \dots \supseteq B_l.
\Eeq
For $j=0,\dots,l$, set
\Beqn DefBUpJ
 B^j(I):= \{f\in C(I,B_j):\ \ f\in C^{j-i}(I,B_i) \ \ \text{for
$i=0,\dots,j$}\}
\Eeq
and equip this space with the corresponding locally convex topology
defined by the seminorms
$
 \|f\|_{B^j([a,b])} :=
 \sum_{i=0}^j \max_{t\in [a,b]} \br\| \frac{d^{j-i}}{dt^{j-i}} f(t)\|_{B_i}
$
where $a,b\in I$, $a<b$. Of course, if $I$ is compact then
$B^j(I)$ is a Banach space.

We call the sequence \eqref{SmSc} a {\em smoothness scale} iff
\begin{itemize}
\item[(i)] $(\A_t)$ descends to a strongly continuous group $(\A_t)$
on each $B_i$;

\item[(ii)] $B_{i+1}$ lies in the domain of the generator of $(\A_t|_{B_i})$;

\item[(iii)]
$\Delta$ restricts to a (necessarily unique) entire superfunction
$\Delta\in\M^{B_i}(\L(B_i))$ for all $i$.
\end{itemize}

We now get a temporal smoothness assertion:

\begin{prp}\label{SelfTmpSm}
Given a smoothness scale \eqref{SmSc},
every solution family $\Xi'\in \M^{B_l(I)}(Z)$
satisfies $\Xi'\in \M^{B^l(I)}(Z)$.
\end{prp}

In case of $B_l$-completeness, it follows that the
universal solution family for the Banach space $B_l$, \ \
$\Xi\sol\in \M^{B_l(\Bbb R)}(\L(B_l)))$, satisfies
$\Xi\sol\in \M^{B^l(\Bbb R)}(\L(B_l)))$,
and hence:

\begin{cor}\label{BUpLSolv}
If the problem \eqref{AbstrIEq} is $B_l$-complete then it is solvable in
$\L(B^l(\Bbb R))$.
\qed\end{cor}

We call an infinite sequence
$B=B_0\supseteq B_1 \supseteq\dots$
an {\em smoothness scale of infinite length} if
for each $l$, the sequence
$B=B_0\supseteq\dots\supseteq B_l$ is a smoothness scale in the sense of
above. Given a smoothness scale of infinite length, set
$B^\infty(\Bbb R):= \bigcap_{i>0} B^i(\Bbb R)$, equipped
with the projective limes topology.
The space of Cauchy data belonging to this is $B_\infty:= \bigcap_{i>0} B_i$,
again with the projective limes topology. Now Cor. \ref{BUpLSolv} and
Cor. \ref{ProjLim} together yield:

\begin{cor}\label{BInfty}
Let be given a smoothness scale of infinite length.
If the problem \eqref{AbstrIEq} is $(B_i,(B_\infty\allowbreak)\seven)$-complete
for $i\gg0$ then it is solvable in $\L(B^\infty(\Bbb R))$.
\qed\end{cor}

\subsection{Support scales}
Here we give an abstract version of causal propagation of perturbations.

A family $(\spsc_t)_{t\in I}$
of closed $\ztwo$-graded subspaces of $B$
where $I\owns0$ is an interval
is called a {\em support scale} if

\begin{itemize}

\item[(ii)] we have $\spsc_t\seq\spsc_{t'}$ for
$0\le t\le t'$ or $t'\le t\le0$;

\item[(ii)] the free evolution "stays within the scale":
If $\xi\in\spsc_0$ then $\A_t\xi\in\spsc_t$ for all $t\in I$;

\item[(iii)] the interaction "is local": For all $t\in I$,
$\Delta\in\M^{B}(\L(B))$ restricts to a (necessarily unique) superfunction
$\Delta\in\M^{B/\spsc_t}(\L(B/\spsc_t))$, i.~e. we have a commutative
diagram
\Bcd
 \L(B) @>\Delta>> \L(B)\nt
 @V VV        @V VV \nt
 \L(B/\spsc_t) @>\Delta>> \L(B/\spsc_t);
\Ecd

\item[(iv)] "Splitting property": There exists a
strongly continuous family $(\E_t)_{t\in I}$ of operators $\E_t:B\to B$
such that for all $\xi\in B$ and $t\in I$ we have
$\E_t\xi-\xi\in\spsc_t$ and
$\br\|\E_t\xi\|_B \le C(t) \br\|\xi\|_{B/\spsc_t}$ with some constant
$C(t)>0$ which is bounded on bounded intervals.
(Thus,  $\E_t$ factors to a bounded operator $B/\spsc_t\to B$ which is a right
inverse to the projection  $B\to B/\spsc_t$.)
\end{itemize}

For two superfunctions $K,K'\in\O^B(Z)$, we will write for shortness
$K \equiv_t K'$ iff $K-K'\in\O^{\spsc_t}(Z)$.

We call a $Z$-family $\Xi'\in\M^{B(I)}(Z)$ of configurations
a {\em relative solution family} (with respect to the support
scale $(\spsc_t)_{t\in I}$) iff
\Beq
 \Xi'(t) \equiv_t \A_t\Xi'(0) + \int_0^t ds \A_{t-s}\Delta[\Xi'(s)]
\Eeq
for all $t\in I$.
With this notion, we get a refined Uniqueness Theorem:

\begin{thm}\label{CausUniqThm}
Fix the problem \eqref{AbstrIEq} and a support scale $(\spsc_t)_{t\in I}$.
Let be given two $Z$-families $\Xi',\Xi"\in\M^{B(I)}(Z)$
which are both relative solution families, and suppose that
$\Xi'(0) \equiv_0 \Xi"(0)$.
Then $\Xi'(t) \equiv_t \Xi"(t)$ for all $t\in I$.
%
\end{thm}

Now Thm. \ref{UniqCor} follows by taking here the trivial support scale
$\spsc_t:=0$ for all $t$.

\subsection{Variants and generalizations}

\subsubsection{Time-dependent interaction}

Consider the problem
\Beqn TDepAbstrIEq
 \Xi'(t) = \A_t\Xi'(0) + \int_0^t ds \A_{t-s}\Delta_s[\Xi'(s)]
\Eeq
where $\Delta_s\in\M^B(\L(B))$ for each $s\in\mathbb R$.

An obvious idea is the reduction onto the time-independent form
\eqref{AbstrIEq} by passing to the enlarged Banach space
$B\ext:=B\oplus\Bbb R\oplus\Bbb R$ and
forming a new one parameter group $(\A_t\ext)$ in $B\ext$
which acts as $\A_t$ in $B$ and as
 $\left(\begin{smallmatrix}
 1 & t \\
 0 & 1
 \end{smallmatrix}\right)$ on
$\Bbb R\oplus\Bbb R$. Setting also
\Beq
 \Delta\ext[\Xi\Cau,\theta,\nu]:=
 \left(\Delta_\theta[\Xi\Cau], 0, 0 \right)
 \in\M^{B\ext}(\L(B\ext)_{\Xi\Cau,\theta,\nu}),
\Eeq
the original problem becomes equivalent with the problem
\Beq
 \Xiexbar(t) = \A_t\ext\Xiexbar(0)
 + \int_0^t ds \A_{t-s}\ext\Delta\ext[\Xiexbar(s)],
\Eeq
together with the initial conditions $\Xiexbar(0)=(\Xi'(0),0,1)$.
Indeed, these enforce every solution family to have the form
$\Xiexbar =(\Xi',t,1)$.

However, this reduction works only if $\Delta_s$ depends real-analytically
on $s$, which makes it unapplicable in the classical field models of
quantum field theory for constraining the interaction onto a finite
space-time domain with the aid of a buffer function $g$ ("adiabatically
switching the interaction").

It is a better idea to generalize the theory by considering an
interaction term to be given as an entire superfunction
$\Delta\in\M^{B(I)}(\L(B))$ the Taylor expansion of which at the origin
has lower degree $\ge2$. Let $\delta_s$ denote evaluation at $s\in I$, and
$\Delta_s:=\delta_s\Delta \in\M^B(\L(B))$. Then
the equation \eqref{TDepAbstrIEq} makes sense for each $s$.
Moreover, it is not hard to show that if leaving $s$ unfixed, the
r.~h.~s. defines an element of $\M^{B(I)}(Z)$. This follows from
the following general fact:

\begin{lem}\label{DeltaLem}
Let $M$ be a finite-dimensional smooth (non-super) manifold,
and $B$ a $\Bbb Z_2$-graded Banach space. For each finite $l\ge0$,
equip $C^l(M,B)$ with the topology of convergence
of derivatives up to $l$-th order on compacta. Also, equip
$C^\infty(M,B)=\bigcap_l C^l(M,B)$ with the projective limes topology.

Let be given a superfunction $\Delta\in\M^{C^l(M,B)}(\L(B))$ where $0\le
j\le\infty$.
Then there exists a unique superfunction
$\Delta \in \M^{C^l(M,B)}(\L(C^l(M,B)))$ (by abuse of notation) which
makes the diagram
\Bcd
 \L(C^l(M,B))  @>\Delta>> \L(C^l(M,B))\nt
 @V\delta_t VV        @VV \delta_t V \nt
 \L(B) @>\Delta_t>> \L(B)
\Ecd
commutative where $\delta_t$ denotes evaluation at $t\in M$, and
$\Delta_t:=\delta_t\Delta \in\M^B(\L(B))$.
\end{lem}

With obvious modifications, our notions and results now carry over to
problems of the form \eqref{TDepAbstrIEq}. In particular,
in the definition of a smoothness scale, condition (iii) has to be
replaced by a condition on temporal smoothness of $\Delta$:
\begin{itemize}
\item[(iii)']
$\Delta\in\M^{B(I)}(\L(B))$
restricts to an entire superfunction
$\Delta\in\M^{B^i(I)}(\L(B_i))$ for all $i$.
\end{itemize}
In adapting the proof of Prop. \ref{SelfTmpSm}, one uses
Lemma \ref{DeltaLem} with $l>0$.


\subsubsection{Source terms}
Another generalization arises by allowing source terms in
\eqref{AbstrDEq}:
\Beq
 \frac d{dt} \Xi'=K\Xi' + \Delta[\Xi'] + J',
\Eeq
or, in integral form,
\Beqn NIEqWSrces
 \Xi'(t) = \A_t\Xi'(0) + \int_0^t ds \A_{t-s}(\Delta[\Xi'(s)]+J'(s)).
\Eeq
We suppose the source term $J'$ to be given as superfunction on a parameter
smf $S$ (this allows sources also for the anticommuting degrees of freedom).

Thus, a senseful Cauchy problem for \eqref{NIEqWSrces}
is to look for $\Xi'\in\M^{B(I)}(Z\times S)$
with given Cauchy data $\Xi'(0)\in \M^B(Z\times S)$ and given source
$J'\in\M^{B(I)}(S)$ which satisfies \eqref{NIEqWSrces} within
$\M^B(Z\times S)$ for $t\in I$.

For technical simplification, we may assume the source to take values in the
Banach space
$B_b(\mathbb R)$ of bounded continuous functions $\mathbb R\to B$ equipped with
the sup norm.

Now there is a universal formulation for this problem which includes
all possible Cauchy data and all possible sources: given
$(\A_t)$ and $\Delta$, we have to find a superfunction
$\Xi\sol\in\M^{B(I)}(\L(B)\times\L(B_b(\mathbb R)))$ such that
\Beqn NUnivWithSources
 \Xi\sol(t) = \A_t\Xi\Cau(0)
 + \int_0^t ds \A_{t-s}(\Delta[\Xi\sol(s)]+J(s))
\Eeq
where $\Xi\Cau,J$ are the standard coordinates on the factors.  

This problem is easily reduced to our standard form \eqref{AbstrIEq}:
we form a new one parameter group $(\A_t\ext)$ in the enlarged Banach space
$B\ext:=B\oplus B_b(\mathbb R)$,
\Beq
 \A_t\ext (\xi\Cau, j) :=
  (\A_t\xi\Cau +  \int_0^t ds \A_{t-s} j(s),\ j(t+\cdot)).
\Eeq
Setting also
$\Delta\ext[\Xi\Cau,J]:=\left(\Delta[\Xi\Cau],
0\right)\in\M^{B\ext}(\L(B\ext)_{\Xi\Cau,J})$,
the problem \eqref{NUnivWithSources} becomes equivalent with the problem
\Beq
 \Xiexbar(t) = \A_t\ext(\Xi\Cau,J)
 + \int_0^t ds \A_{t-s}\ext\Delta\ext[\Xiexbar(s)],
\Eeq
which has our standard form \eqref{AbstrIEq}.

With a similar trick, one can also treat non-dynamical fields.

\subsubsection{Semigroups}
An obvious way to generalize \eqref{AbstrIEq} is to replace the
strongly continuous group $(\A_t)_{t\in\Bbb R}$ on $B$ by a
strongly continuous semigroup $(\A_t)_{t\ge0}$. In that case, only
configuration families $\Xi'\in\M^{B(I)}(Z)$ with $I\seq\Bbb R_+:=\{t\ge0\}$
are to be taken into account.

All our results generalize mutatis mutandis onto this case; if the problem is
complete we get a universal solution supermanifold
$\L(B(\Bbb R_+))\sol \seq\L(B(\Bbb R_+))$.
(Note, however, that anticommuting degrees of freedom occur mainly in classical
field
models of quantum field theory, where the time evolution is always
time-reversible.)

\subsubsection{Non-entire interaction: Cauchy uniqueness}

The reader will note that in the original problem \eqref{AbstrIEq}, the
entireness hypothesis on $\Delta$  will be not needed for showing Cauchy
uniqueness;
it will be used only for the construction of the short-time solution.

In order to formulate Cauchy uniqueness in its most general form, we go a step
further and consider a generalization of the problem \eqref{AbstrIEq}
by supposing only $\Delta=\Delta[\Xi]\in \M^B(U)$ where $U\seq\L(B)$ is an open
subset of $B\seven$ which contains $0$, considered as sub-superdomain.
However, we keep the requirement that the Taylor expansion of $\Delta$
in zero has lower degree $\ge2$.

The notion of a configuration family has to be modified:
we require additionally that for the underlying function
$\widetilde{\Xi'}: \opn space (Z)\to B(I)\seven$ of $\Xi'\in\M^{B(I)}(Z)$,
we have $\widetilde{\Xi'}(t)\in\opn space (U)$ for all $t\in I$.

For any compact interval $I\owns0$, set $U(I):= C(I,U)$; this is open in
$B(I)\seven$,
and hence is the
underlying space of an open sub-superdomain in $\L(B(I))$ which we abusively
denote by $\L(U(I))$. Now a configuration family is the same as a morphism
$\Xi':Z\to\L(U(I))$.

For such a  configuration family, the r.~h.~s. of \eqref{AbstrIEq} is now
well-defined. Of course,
we call $\Xi'$ again a {\em solution family} iff \eqref{AbstrIEq} holds.

The Cauchy uniqueness still generalizes to this situation. The proof
of Thm. \ref{UniqCor} actually yields:

\begin{cor}
Fix the problem \eqref{AbstrIEq} where $B$ and $(\A_t)_{t\ge0}$
are as in \ref{ConfFam},
and $\Delta=\Delta[\Xi]\in\M^B(U)$, $U\seq\L(B)$ open.
Suppose that $0\in U$, and that the Taylor expansion $\Delta_0$
has lower degree $\ge2$.
Given solution families $\Xi',\Xi": Z\to\L(U(I))$ where $I\owns0$ is connected
such that for some $t_0\in I$ we have $\Xi'(t_0)=\Xi"(t_0)$, we have
$\Xi'=\Xi"$.
\qed\end{cor}

\subsubsection{Non-entire interaction: Short-time existence }
For $\Delta$ defined only on some open $U$, looking for all-time existence
is not very senseful. However, the approach to short-time existence
given in Prop. \ref{ShrtTime} below generalizes: the assertion (i) on the
existence
of a formal solution remains unchanged (it only uses
the formal power series $\Delta_0$),
while for analyticity we have to make a certain trade-off in the domain of
definition (which is clearly necessary since the free evolution
has to stay at least for a short time in the domain of definition of the
interaction):

\begin{cor}
Let $U'\in\CB (B)$ such that $\Delta_0\in\P(B,U';B)$. For each
$c< 1/ \limsup_{t\to0}\br\|\A_t\|$ there exists $\theta$ such that
$\Xi\sol\in\P(B,cU';B([-\theta,\theta]))$.
\qed\end{cor}


\subsubsection{Grassmann-valued solutions}\label{AbstSolValGrass}
The most naive notion of a configuration in a classical field model with
anticommuting fields arises by replacing the domain $\Bbb R$ for the real
field components by a finite-dimensional Grassmann algebra
$\Lambda_n=\Bbb C[\zeta_1,\dots,\zeta_n]$ (we recall that, in accordance with
our hermitian framework, only complex Grassmann algebras should be used).
Thus, a {\em $\Lambda_n$-valued configuration}
is an element $\xi\in \left(\Lambda_n\otimes B(I)\right)_{\even,\Bbb R}$.

Now denote by $Z_n$ the unique connected $0|n$-dimensional smf, which is just a
point together with the Grassmann algebra $\O(Z_n)=\Lambda_n$. Because of
$\Lambda_n\otimes B(I)=\O^{B(I)} (Z_n)$, such an element $\xi$ is the
same as a  $Z_n$-family.
Also, $\xi$ is a solution family in our sense iff the equation
\eqref{AbstrIEq} is satisfied within $\Lambda_n\otimes B$.
We now get an overview over all $\Lambda_n$-valued solutions:

\begin{cor}
Suppose that the problem \eqref{AbstrIEq} is complete, and let be given
$\Lambda_n$-valued Cauchy data
$\xi\Cau\in \left(\Lambda_n\otimes B\right)_{\even,\Bbb R}$.
Then there exists a unique solution $\xi$
with these Cauchy data. It is given by
\Beq
 \xi = \Xi\sol[\xi\Cau] =\Xi\sol_{b(\phi\Cau)}[s(\xi\Cau)]
\Eeq
where $b(\cdot): \Lambda_n\to\Bbb C$ denotes the body map,
and $s(\cdot)= 1-b(\cdot)$ the soul map.
\qed\end{cor}

(For a discussion in the context of evolution PDEs as well as
of solutions in the infinite-dimensional Grassmann algebra
$\Lambda_\infty$ of supernumbers introduced by deWitt \cite{[DeWitt]}, cf.
\cite{[CAUCHY]}.)


\section{Application to systems of evolution equations}

\subsection{The setting}
Here we fix a class of systems of classical
nonlinear wave equations in Minkowski space $\rdmm$ which is wide enough to
describe the field equations of many usual models, like e.~g. $\Phi^4$,
quantum electrodynamics, Yang-Mills theory with usual
gauge-breaking term, Faddeev-Popov ghosts, and possibly
minimally coupled fermionic matter. The novelty in our equations is the
appearance of anticommuting fields; in describing the system, they simply
appear as anticommuting variables generating a differential power series
algebra. However, it is no longer obvious what a solution of our system
should be. In fact, as argued in \cite{[WHAT]}, there are no longer
"individual" solutions (besides purely bosonic ones, with all fermionic
components put to zero); but it is sensible to look for {\em families} of
solutions parametrized by supermanifolds. In particular, solutions with
values in Grassmann algebras can be reinterpreted as such families
(cf. \ref{AbstSolValGrass} and \cite{[CAUCHY]}).

We will consider the system of partial differential equations
in $\rdmm$
\Beqn TheSyst
 L_i[\Xi'] \equiv \partial_t \Xi'_i - \sum_{j=1}^N K_{ij}(\partial_x)\Xi'_j
 - \Delta_i[\Xi'] =0 \quad (i=1,\dots, N=N\seven+N\sodd).
\Eeq
Here $\Xi'=(\Xi'_1,\dots,\Xi'_N)
 =(\Phi'_1,\dots,\Phi'_{N\seven}|\Psi'_1,\dots,\Psi'_{N\sodd})$ is a tuple of
$N\seven$ commuting, ordinary, "bosonic" fields as well as of
$N\sodd$ anticommuting, "fermionic" fields.
The {\em kinetic operator} $K_{ij}(\partial_x)$ is a real differential
operator with constant coefficients and containing only spatial derivatives. We
demand that
parities are preserved, i.~e. $K_{ij}(\partial_x)=0$ if
$\br|\Xi_i|\not=\br|\Xi_j|$; additional
requirements will be specified below.

The {\em interaction terms} $\Delta_i[\ul\Xi]$ are real, entire
differential power series (in the finite-dimensional sense)
of lower degree $\ge 2$, i.~e.
\Beq
 \Delta_i[\ul\Xi]= \Delta_i[\ul\Phi|\ul\Psi]\in
 \Bbb C[[(\partial^\nu\ul\Xi_i
  )_{i=1,\dots,N,\ \nu\in\Bbb Z_+^d,\ \br|\nu|\le n}]]
\Eeq
for some $n\ge0$, where, as usual,
$\partial^\nu:= \partial_1^{\nu_1}\cdots\partial_d^{\nu_d}$.
As in \cite{[CMP1]}, the underlined letters $\ul\Phi,\ul\Psi$ denote
the even and odd indeterminates of an algebra of differential polynomials or
differential power series, while the non-underlined letters
$\Xi,\Phi,\Psi$ denote superfunctions or their Taylor expansions.
(As usual, a power series in a finite number of even and odd variables,
$P[y|\eta]=\sum P_{\mu\nu}y^\mu\eta^\nu \in
 \Bbb C[[y_1,\dots,y_m|\eta_1,\dots,\eta_n]]$
is entire iff for all $R>0$ there exists $C>0$  such that
$\br|P_{\mu\nu}| \le C R^{-|\mu|}$ for all $\mu,\nu$.)
Of course, we also require that $\Delta_i$ is even for
$i=1,\dots,N\seven$ and odd for $i=N\seven+1,\dots,N\seven+N\sodd$.

We require that there exist  integers
$\sd_1,\dots,\sd_N$, called {\em smoothness offsets}, with the following
properties:

I. There exist $t_0>0$,  $C>0$ such that the matrix-valued function
\Beqn DefHatA
 \hat A: \Bbb R\times\Bbb R^d\to\Bbb C^{N\times N},\quad
 \hat A(t,p) :=  (2\pi)^{-d/2} \exp( K(\i p) t)1_{N\times N}
\Eeq
satisfies the estimate
\Beqn TheSmCEst
 \br\| \hat A_{ij}(t,p) \| \le C(1+\br|p|)^{\sd_j-\sd_i},
\Eeq
for $p\in\rdm$, $t\in[-t_0,t_0]$ with suitable $t_0>0$,\ $C>0$.

II. For all $i,k=1,\dots,N$, $\nu\in{\Bbb Z_+}^n$, we have
\Beqn SmCond
 \frac\partial{\partial(\partial^\nu\ul\Xi_k)} \Delta_i[\ul\Xi]\ne0
 \qquad
 \Longrightarrow
 \qquad
 \max(0,\ \sd_i) \le \sd_k-\br|\nu|.
\Eeq

\Brm
(1) The function $\hat A$ satisfies the spatially
Fourier-transformed and complexified free field
equations,
 $\frac d{dt} \hat A (t,p) - K(\i p) \hat A = 0$,\ \ 
 $\hat A(0,p)= (2\pi)^{-d/2}1_{N\times N}$.
(Our convention for Fourier transforms is $\hat f(p) = {\mathcal F}_{x\to
p}f(p)
 = (2\pi)^{-d/2}\int_{\rdm} dx \e^{-\i px}f(x)$
for $f\in {\mathcal S}(\rdm)$.)

(2) Obviously, the estimate \eqref{TheSmCEst} implies hyperbolicity of the
kinetic operators, i. e. for all $p\in\rdm$, the matrix $K(\i p)$ has only
imaginary eigenvalues.

(3) Usually, the smoothness offsets save that smoothness information which
would be otherwise lost in reducing a temporally higher-order system to a
temporally first-order one.

(4) The smoothness condition \eqref{SmCond} is rather constraining;
it excludes e. g. the Korteweg-de Vries equation as well as the nonlinear
Schr\"odinger
equations. Fortunately, it is satisfied for apparently all wave equations
occurring in quantum-field theoretical models. (Of course, the smoothness
offsets
have to be chosen suitably: usually, one for second-order fields, and zero for
their derivatives as well as for first-order fields.)

(5) In \cite{[CAUCHY]},  we had constrained the smoothness offsets to be
nonnegative.
\Erm

\subsection{Basic results}
We use the standard Sobolev spaces:
For real $k>d/2$, let $H_k(\rdm)$ be the space of all
$f\in L_2(\rdm)$ for which $(1+ \br|p|)^k\hat f(p)$ is square-integrable.
Our basic Banach space of Cauchy data is
\Beq
 \cH\V_k:= \bigoplus_{i=1}^{N\seven} H_{k+\sd_i}(\rdm) \quad\oplus\quad
 \bigoplus_{i=1}^{N\sodd} \Pi H_{k+\sd_i}(\rdm).
\Eeq
Because of \eqref{TheSmCEst} we can take the inverse spatial Fourier transform
$A(t,x)$ of the function $\hat A(t,p)$ defined in \eqref{DefHatA}, and it
follows that $K(\partial_x)$ is the generator of the continuous
one-parameter group $(\A_t)$ in $\cH\V_k$ given by
\Beqn Concr1PG
 \A_t\xi(x):= \int_{\rdm} dy A(t,x-y)\xi(y),\quad\text{i.~e.\ \ }
\widehat{\A_t\xi}(p):= (2\pi)^{d/2}\hat A(t,p)\hat \xi(p).
\Eeq
In order to assign to the $\Delta_i$ an entire superfunction $\Delta[\Xi\Cau]$,
we split them by degree:
$\Delta_i =\sum_{l\ge2} \Delta_{i,(l)}$.
Thus, $\Delta_{i,(l)}[\ul\Xi]$ is a differential polynomial, and,
due to the condition \eqref{SmCond}, the substitution
$\partial^\nu\ul\Xi_i\mapsto\partial^\nu\Xi\Cau_i$ yields a polynomial
superfunction $\Delta_{i,(l)}[\Xi\Cau]\in\O^{H_{k+\sd_i}(\rdm)}(\L(\cH\V_k))$.
On the other hand, we have a Fr\`echet topology
on the subspace
$\O^{H_{k+\sd_i}(\rdm)}(\L(\cH\V_k))_{\opn ent }$
of entire superfunctions (cf. \ref{InfDimSGeom}) by the seminorms
\Beq
 f\mapsto \br\|f_0\|_{\P(\cH\V_k,nU;H_{k+\sd_i}(\rdm))}
\Eeq
where $n=1,2,\dots$,\ \  $f_0$ is the Taylor expansion at zero,
and $U$ is the unit ball. In this topology,
the series $\Delta_i[\Xi\Cau] :=\sum_{l\ge2} \Delta_{i,(l)}[\Xi\Cau]$
converges, and hence
\[
 \Delta[\Xi\Cau]:= (\Delta_1[\Xi\Cau],\dots,\Delta_{N\seven}[\Xi\Cau],
 \Pi\Delta_{N\seven+1}[\Xi\Cau],\dots,\Pi\Delta_{N\seven+N\sodd}[\Xi\Cau])
 \in\M^{\cH\V_k}(\L(\cH\V_k))
\]
is a well-defined entire superfunction. Thus, we can rewrite \eqref{TheSyst}
into integral form:
\Beqn TheIntSystem
 \Xi'(t,x) = \int_{\rdm} dy A(t,x-y)\Xi'(0,y) +
 \int_0^t ds \int_{\rdm} dy A(t-s,x-y) \Delta[\Xi'(s,\cdot)](y).
\Eeq
This has the form of the abstract problem \eqref{AbstrIEq} with
$B:=\cH\V_k$, and $\A_t$ being given by \eqref{Concr1PG}.
A superfunction $\Xi'\in\M^{\cH\V_k(I)}(Z)$
is a solution family if and only if it satisfies \eqref{TheSyst}
within $\M^{\cD'(I\times\rdm)}(Z)$.
On the other hand, an element
$\phi=(\phi_1,\dots,\phi_{N\seven})\in\cH\V_k(I)\seven$
satisfies the underlying system \eqref{UnderlEvProbl} iff the functions
$\phi_i\in C(I,\ H_{k+\sd_i}(\rdm))$ fulfill
\Beqn UnderlSyst
 \partial_t \phi_i - \sum_{j=1}^{N\seven} K_{ij}(\partial_x)\phi_j
 - \Delta_i[\phi|0] =0 \quad (i=1,\dots, N\seven).
\Eeq
Note that $\phi\mapsto\Delta_i[\phi|0]=\tilde\Delta_i[\phi]$ is the underlying
function of
the superfunction $\Delta$. Now Thm. \ref{ComplThm} specializes to:

\begin{cor}\label{ComplSystCor}
Fix some $k>d/2$.
For a subset $A\seq(\cH\V_k)\seven$, the following conditions are equivalent:

(i) For every solution $\phi\in \cH\V_k((a,b))\seven$ of
the underlying system \eqref{UnderlSyst} on a bounded open time
interval $(a,b)\owns 0$ such that $\phi(0)\in A$, we have
\Beq
\sup_{t\in(a,b)} \br\|\phi(t)\|_{\cH\V_k} < \infty.
\Eeq

(ii) The underlying system \eqref{UnderlSyst} is all-time solvable for
Cauchy data in $A$.

(iii) Whenever we are given an smf $Z$ and a superfunction
$\XiCauP\in\M^{\cH\V_k}(Z)$ such that the image of the underlying function
$\widetilde{\XiCauP}:\opn space (Z)\to (\cH\V_k)\seven$
is contained in $A$, there exists a (necessarily uniquely determined)
solution family $\Xi'\in\M^{\cH\V_k(\Bbb R)}(Z)$ of \eqref{TheSyst}
with $\Xi'(0) = \XiCauP$.

\noindent If these conditions are satisfied we call the the problem
\eqref{TheSyst} {\em $(\cH\V_k,A)$-complete}. If it is
$(\cH\V_k,(\cH\V_k)\seven)$-complete we call it simply {\em
$\cH\V_k$-complete}.
\qed\end{cor}

It follows from Cor. \ref{CorSolvInLBR} that if the problem \eqref{TheSyst} is
$\cH\V_k$-complete it defines
an smf of classical solutions $\L(\cH\V_k(\Bbb R))\sol \allowbreak\seq
\L(\cH\V_k(\Bbb R))$.

\begin{prp}\label{SobSlfSm}
If the problem \eqref{TheSyst} is $(\cH\V_k,A)$-complete with a subset
$A\seq(\cH\V_{k+l})\seven$ where $l>0$ is integer
then it is $(\cH\V_{k+l},A)$-complete.
\end{prp}

Set
\Beq
 \smloss:= \max \br\{1, \max_{i,j=1,\dots,N}
  \br({\sd_i-\sd_j+\opn ord K_{ij}(\partial_x)})\}
\Eeq
where $\opn ord K_{ij}(\partial_x)$ is the order of the differential operator
($=-\infty$ if $K_{ij}=0$). Then, fixing $k>d/2$, the sequence
$\cH\V_{k}\supseteq\cH\V_{k+\mu}\supseteq\cH\V_{k+2\mu}\supseteq\dots$
forms an infinite smoothness scale, and
Prop. \ref{SelfTmpSm} and Cor. \ref{BUpLSolv} apply.

Set $\cH\V_\infty:=\bigcap_{k>d/2} \cH\V_k$. This is the
space of Cauchy data belonging to $C^\infty(\Bbb R,\cH\V_\infty)$, and
Cor. \ref{BInfty} yields:

\begin{cor}
If the system \eqref{TheSyst} is $(\cH\V_k,(\cH\V_\infty)\seven)$-complete
for some $k>d/2$ then it is solvable in $\L(C^\infty(\Bbb R,\cH\V_\infty))$.
\qed\end{cor}

The space $\cH\V_\infty$ lies between the Schwartz space
$\mathcal S(\rdm) \otimes\Bbb R^{N\seven|N\sodd}$
and $C^\infty(\rdm)\otimes\Bbb R^{N\seven|N\sodd}$.
It would be interesting to know how to descend to the Schwartz space.

\subsection{Causality}
In this section, we study the consequences of finite propagation speed, as
it holds in classical field theories used in quantum field theory.

For $(s,x),(t,y)\in\rdmm$ we will write $(s,x)\causall(t,y)$ iff $(t,y)$ lies
in the forward light cone of $(s,x)$, i.~e. $\br|t-s|\ge\br|y-x|$.
We call the system \eqref{TheSyst} {\em causal} iff we have
(cf. \eqref{Concr1PG})
\Beq
   \supp A \seq \{(t,x)\in\rdmm:\quad \br|x|\le \br|t|\}.
\Eeq
Given a point $p=(s,x)\in\rdmm$ with $s\not=0$,  write
\Bal
 &\Omega(p) :=
 \begin{cases}
  \{(s',x')\in\rdmm:\ (s',x') \causall (s,x),\ 0 < s'\}  & \text{if $s>0$,}\\
  \{(s',x')\in\rdmm:\ (s,x) \causall (s',x'),\ s' < 0\}  & \text{if $s<0$,}
 \end{cases}
 \\
 &\J(p) := \{x'\in\rdm:\ \br|x'-x| < \br|s|\}.
\Eal
As to be expected, causality implies that perturbations of solution families
propagate within
the light cone:

\begin{thm}\label{CausCauUniqu}
Suppose that the system is causal and $k>d/2$.

(i) Let be given a point $p=(s,x)\in\rdmm$ with $s\not=0$,
and let $I=[0,s]$ if $s>0$ and $I=[s,0]$ if $s<0$, respectively.

Let be given two $Z$-families $\Xi',\Xi"\in\M^{\cH\V_k(I)}(Z)$, and suppose
that
\Bea
 &L_i[\Xi']|_{\Omega(p)} = L_i[\Xi"]|_{\Omega(p)} =0
 \quad (i=1,\dots, N),
 \\
 &\left(\Xi'(0)-\Xi"(0)\right)|_{\J(p)}=0.
\Eea
Then $\br(\Xi'-\Xi")|_{\Omega(p)}=0$.

(ii) Suppose that for a solution family $\Xi'\in\M^{\cH\V_k(I)}(Z)$ with
$I\owns0$ satisfies $\Xi'(0,x)=0$ for $\br|x| > r$
with some $r>0$. Then $\Xi'(t,x)=0$ for $\br|x| > \br|t|+r$ and $t\in I$.
\end{thm}

For $r\ge0$, let
\Beqn bVr
 \bV_r:=\{(t,x)\in\rdmm:\quad \br|x|\le r+\br|t|\},
\Eeq
and set
\Beqn SpcEc
 C^\infty_c(\rdmm,\Bbb R^{N\seven|N\sodd}) = \bigcup_{r>0}
 \{f\in C^\infty(\rdmm,\Bbb R^{N\seven|N\sodd}):\ \ \supp f \seq \bV_r\}.
\Eeq
Equipping each item of the union with the closed subspace topology and
\eqref{SpcEc} with the arising inductive limit topology, this is a strict
inductive limes of Fr\`echet spaces, and hence complete. Also, $\cD(\rdmm)$
is dense in \eqref{SpcEc}; hence \eqref{SpcEc} is admissible in the sense
of \CMPDefAdm. Moreover, it is important for field-theoretical applications
that the Poincar\'e group acts continuously on \eqref{SpcEc}.
Of course, the space of Cauchy data belonging to \eqref{SpcEc} is the
testfunction
space $\cD(\rdm,\Bbb R^{N\seven|N\sodd})$.

Our main result for the causal case is:

\begin{thm}\label{cDMainThm}
If the system \eqref{TheSyst} is both causal and
$(\cH\V_k,\cD(\rdm,\Bbb R^{N\seven|0}))$-complete for some $k>d/2$,
then it is solvable in $\L(C^\infty_c(\rdmm,\Bbb R^{N\seven|N\sodd}))$.
\end{thm}


We want to show also solvability in smooth functions,
\Beq
 \Ci:=C^\infty(\rdmm,\Bbb R^{N\seven|N\sodd}).
\Eeq
However, this does not quite fit into our general scheme since there is no
Banach space $B$ of functions on $\rdm$ such that
$\Ci\seq B(\Bbb R)$ (indeed, there is no continuous norm on $\Ci$). Therefore
we note
that if the system \eqref{TheSyst} is causal then
for $\Xi'\in \M^{\Ci}(Z)$,
both the system \eqref{TheSyst} and the integrated version
\eqref{TheIntSystem} make sense and are equivalent; if they are satisfied
we call $\Xi'$ a {\em smooth solution family}.
Of course, the appropriate space of Cauchy data is
$\CiCau:=C^\infty(\rdm,\Bbb R^{N\seven|N\sodd})$.

\begin{thm}\label{MainThmSm}
If the system \eqref{TheSyst} is both causal and
$(\cH\V_k,\cD(\rdm,\Bbb R^{N\seven|0}))$-complete for some $k>d/2$,
then it is solvable in $\L(\Ci)$ in the following sense:

There exists a (necessarily uniquely determined) superfunction
$\Xi\sol\in\M^{\Ci}(\L(\CiCau))$
such that $\Xi\sol$ is a smooth solution family, and
$\Xi\sol(0) = \Xi\Cau$ where $\Xi\Cau\in \M^{\CiCau}(\L(\CiCau))$ is
the standard coordinate. Moreover,
the image of morphism $\Xi\sol: \L(\CiCau)\to\L(\Ci)$ is a
split sub-smf which we call the {\em smf of smooth classical solutions}, and
denote by
$\L(\Ci)\sol$.
\end{thm}

Of course, the consequences of solvability are the same as in
\ref{DefSolvable}.
In particular, the underlying manifold $\widetilde{\L(\Ci)\sol}$
identifies with the set of all $\phi\in C^\infty(\rdmm,\Bbb R^{N\seven})$
which satisfy \eqref{UnderlEvProbl}. Also, we
get a commutative diagram
\Bcd
 \L(\cD(\rdm,\Bbb R^{N\seven|N\sodd})) @>\Xi\sol>>
  \L(C^\infty_c(\rdmm,\Bbb R^{N\seven|N\sodd}))
 \nt
 @V VV        @VV V \nt
 \L(\CiCau) @>\Xi\sol>> \L(\Ci).
\Ecd
For a further variant, which considers spatially compactly carried excitations
of solutions, and therefore is interesting in the context of spontaneous 
symmetry breaking, cf. \cite{[CAUCHY]}.

\section{Proofs}\label{Proofs}
\subsection{Short-time results}
We will need the following standard fact on strongly continuous operator
groups:
There exists a constant $C_1>0$ such that we have for $\theta\in(0,1]$
and $\xi\in B$
\Beqn EstA
 \br\|\A_\mutarg\xi]\|_{B([-\theta,\theta])}
  \le C_1\br\|\xi\|_{B}.
\Eeq
It follows that for $\theta\in(0,1]$, $g\in B(\Bbb R)$ we have
\Beqn EstIntAg
 \br\|\int_0^\mutarg ds \A_{\mutarg-s} g(s)\|_{B([-\theta,\theta])}
 \le C_1\theta \br\|g\|_{B([-\theta,\theta])}.
\Eeq
In solving the problem \eqref{AbstrIEq},
we first construct the Taylor expansion at zero of the superfunctional
$\Xi\sol$ sought for; we will denote it by $\Xi\sol$ again.

\begin{prp}\label{ShrtTime}
(i) There exists a uniquely determined formal power series
\Beqn TheFormSol
 \Xi\sol=\Xi\sol[\Xi\Cau]\in\Pf(B;B(\Bbb R))
\Eeq
which solves \eqref{AbstrIEq} within $\Pf(B;B)$. Explicitly, we have
\Beq
 \Xi\sol_{(\le1)}(t)=\A_t\Xi\Cau, \qquad
 \Xi\sol_{(n+1)}(t) = \int_0^t ds \A_{t-s}\Delta[\Xi\sol_{(\le n)}]_{(n+1)}(s)
\Eeq
for $n\ge1$. We call \eqref{TheFormSol} the {\em formal solution}
of the problem \eqref{AbstrIEq}.

(ii) The formal solution is "short-time analytic":
For any $c>0$ there exists $\theta>0$ such that
$\Xi\sol\in\P(B,cU;B([-\theta,\theta]))$ where $U\seq B$ is the unit ball.
\end{prp}

(Of course, in (ii) we have silently applied the restriction map $B(\Bbb R)\to
B([-\theta,\theta])$ in the target.)

\begin{proof}
Ad (i). This follows by splitting \eqref{AbstrIEq} into degrees.

Ad (ii).
For $n\ge0$, we have from \eqref{AbstrIEq}
\Beqn IndKey
 \Xi\sol_{(\le n+1)}(t) = \A_t\Xi\Cau +
 \int_0^t ds \A_{t-s}\Delta[\Xi\sol_{(\le n)}]_{(\le n+1)}(s).
\Eeq
We will show that for sufficiently small $\theta>0$ we have for all
$n\ge0$ the estimate
\Beqn IndAss
 \br\|\Xi\sol_{(\le n)}\|\le 2C_1c
 \quad\text{within $\P(B,cU;B([-\theta,\theta]))$.}
\Eeq
Passing to the limit $n\to\infty$ we get the assertion.

{}From the hypothesis on entireness and the absence of a constant term in
$\Delta$, we have:

\begin{lem}
Given $C'>0$, there exists $C">0$ with the following property:
If $E$ is a $\Bbb Z_2$-lcs, $p\in\CS(E)$  and the power series
$\Xi'\in \P(E,p;B([-\theta,\theta]))$ satisfies  $\br\|\Xi'\|<C'$, then
\Beq
 \br\|\Delta[\Xi']\|\le C"
 \quad\text{within $\P(E,p;B([-\theta,\theta]))$.}
\Eeq
\qed\end{lem}

We now prove \eqref{IndAss} by induction on $n$.
The start of induction, $n=1$, follows from \eqref{EstA}.
Now, for $n\ge1$, we find from
\eqref{IndKey} that within $\P(B,cU;B([-\theta,\theta]))$
\Beq
 \br\|\Xi\sol_{(\le n+1)}\| \le \br\|\A_\mutarg\Xi\Cau\| +
 \br\|\int_0^\mutarg ds
  \A_{\mutarg-s}\left(\Delta[\Xi\sol_{(\le n)}]_{(\le n+1)}\right)\|.
\Eeq
Using \eqref{EstA} and \eqref{EstIntAg}, this becomes
$\le C_1c + C_1\theta\br\|\Delta[\Xi\sol_{(\le n)}]_{(\le n+1)}\|$.
Because of the hypotheses of induction, the preceding Lemma
applies with $C':=2C_1c$, yielding
\Beq
 \br\|\Xi\sol_{(\le n+1)}\| \le C_1c + C_1\theta C",
\Eeq
and the assertion of induction, $\br\|\Xi\sol_{(\le n+1)}\| \le 2C_1c$,
is satisfied for $\theta\le c/C"$. The Proposition is proved.
\end{proof}

The condition \eqref{AbstrIEq} makes also sense if
$\Xi'$ is only a power series $\Xi'\in\P(F;B(I))$
where $F$ is an arbitrary $\Bbb Z_2$-lcs; if it is satisfied we call
$\Xi'$ a {\em solution power series}.

Of course, the Taylor expansions
$\Xi"_z$ ($z\in Z$) of any solution family $\Xi"\in\M^{B(I)}(Z)$ are
solution power series. Conversely, if $I$ is compact then
the target $B(I)$ is a Banach space, too, and hence for any element
$\Xi'\in\P(F;B(I))$ there exists some
$p\in\CS(F)$ such that $\Xi'\in\P(F,p;B(I))$; by \CMPDefinesFGerm, it follows
that
any solution power series defines a solution family
$\Xi'\in\M^{B(I)}(U)$ on the open unit ball $U$ within the superdomain
$\L(\hat F_p)$. Hence we can switch rather freely between solution families
and solution power series.

\subsection{Long-time results}

The existence of long-time solutions for the underlying even problem
implies the existence of long-time solution families for the original
problem:

\begin{lem}\label{XiSolXl}
Let $\phi\in B\seven$ be such that there exists a
solution of the underlying even problem
$\phi'\in B([0,b))\seven$ with $\phi=\phi'(0)$ and
\Beq
 \sup_{t\in[0,b)} \br\|\phi'(t)\|_B <c' < \infty.
\Eeq
Then there exists some $\epsilon>0$ which depends only on $c$ and a
(necessarily uniquely determined)
solution power series
$\XiCData=\XiCData[\Xi\Cau]\in\P(B;B([0,b+\epsilon)))\sevR$
such that $\XiCData(0) = \Xi\Cau+\phi$.
\end{lem}

\Brm
$\XiCData$ will become the Taylor expansion of the superfunction $\Xi\sol$ at
$\phi$,
motivating the notation.
\Erm

\begin{proof}
First we note that there exists a solution power series
$\Xi'\in\P(B;B([0,b')))\sevR$ with some $b'>0$ such that
$\Xi'(0) = \Xi\Cau+\phi$ (indeed, using Prop. \ref{ShrtTime}.(ii) with
$c:=\br\|\phi\|+1$, the translation (cf. \CMPTransl)
$\Xi':=\opn t _{\phi}\br(\Xi\sol|_{[0,\theta)})$ of $\Xi\sol$ by $\phi$
has this property with $b':=\theta$).
By Cauchy uniqueness (cf. Thm. \ref{UniqCor}), such a solution power series
exists either for each $b'$,
or there is a maximal $b'$ such that such a solution power series exists
(roughly spoken, this $b'$ is just the forward lifetime for the Cauchy datum
$\phi$).
If the assertion is wrong then such a maximal $b'$ exists and is $\le b$.
Now, since the absolute term $\Xi'[0]\in B([0,b'))$ is a solution
of the underlying even problem, Thm. \ref{UniqCor} implies
$\Xi'[0]=\phi'|_{[0,b')}$.

By Prop. \ref{ShrtTime}.(ii), there exists
$\theta>0$ such that $\Xi\sol\in\P(B,c'U;B([-\theta,\theta]))$
where $U$ is the unit ball.
Composing $\Xi\sol$ with $\Xi'(b'-\theta/2)\in\P(B;B)\sevR$
yields a solution power series
$\Xi":=\Xi\sol[\Xi'(b'-\theta/2)]\in\P(B;B([-\theta,\theta]))$.
We perform a time shift:
$\Xi"(\mutarg-b'+\theta/2)\in\P(B;B([b'-3\theta/2,b'+\theta/2]))$.
Now $\Xi"(\mutarg-b'+\theta/2)$ and $\Xi'$, being solution power series with
the same
Cauchy data at time $b'-\theta/2$, join together to a
solution power series $\XiCData\in\P(B;B([0,b'+\theta/2)))$ which extends
$\Xi'$, in contradiction to our assumption.
\end{proof}

\begin{proof}[Proof of Thm. \ref{ComplThm}]
(iii)$\Rightarrow$(ii) is obvious, and (ii)$\Rightarrow$(i) is clear from Thm.
\ref{UniqCor} applied to
$Z$ being a point. For (i)$\Rightarrow$(ii), one uses the preceding Lemma.
Turning to (ii)$\Rightarrow$(iii), we may assume that $Z\seq\L(F)$ is a
superdomain. We will
show that the assignment
\Beq
 \opn space (Z)\owns z\mapsto \Xi'_z:=\Xi\sol_{\lambda(z)}\in\P(F;B(\Bbb R))
\Eeq
where $\lambda:=\widetilde{\XiCauP}$, and $\Xi\sol_{\lambda(z)}$ is defined
by Lemma \ref{XiSolXl},
is a superfunction $\Xi'\in\M^{B(\Bbb R)}(Z)$.

Recalling the definition of the topology of $B(\Bbb R)$, it is sufficient
to show that for $a<0<b$, the assignment
$z\mapsto\Xi'_z|_{[a,b]}$ is an
element of $\M^{B([a,b])}(Z)$. Indeed, fix $z$, and choose
a continuous seminorm $p$ on $F$ with $\Xi'_z|_{[a,b]}\in\P(F,p;B([a,b]))$.
For $z'\in F\seven$ with $p(z')<1$,\ \   $\Xi'_{z+z'}|_{[a,b]}$ is a
solution power series with the same Cauchy data
as the translation $\opn t _{z'} \Xi'_z|_{[a,b]}$, hence they coincide,
proving our assertion.
\end{proof}

We remark that without any completeness hypothesis, one gets from
Prop. \ref{ShrtTime}.(ii) at least an existence result for short-time
solution families:

\begin{cor}
Let be given an smf $Z$ and an smf morphism $\XiCauP: Z\to\L(B)$.
Suppose that the image of the underlying map
$\widetilde{\XiCauP}:\opn space (Z)\to B\seven$ is bounded. Then
there exists an open interval $I\owns0$ and a solution family
$\Xi'\in\M^{B(I)}(Z)$ with $\Xi'(0) = \XiCauP$.
\qed\end{cor}

The following (more or less standard) method allows to conclude from
$B$-completeness
to $B'$-completeness where $B'$ is a "smaller" Banach space.

\begin{lem}\label{SelfSm}
(i) Let be given continuous seminorms $p,q\in\CS(B)$ with $p\le q$, and assume
that there exists a function $K(\cdot)$ which is bounded on finite
intervals such that $q(\A_t(b)) \le K(t)q(b)$ for all $b\in B\seven$.
Let be given a solution $\phi\in B'(I)\seven$ of the underlying problem with
$I=(t_1,t_2)\owns0$, such that both $p(\phi(t))$ and $p(\tilde\Delta(\phi(t)))$
are bounded
(we recall that $\tilde\Delta$ is the underlying function of the superfunction
$\Delta$).
Suppose that there exists a function $C(\cdot)$ which is bounded on finite
intervals such that
\Beq
  q(\tilde\Delta(b)) \le C(p(b))(1+q(b))
\Eeq
for $b\in \opn Im \phi$. Then $q(\phi(t))$ will be bounded, too.

(ii) Let be given a continuous, even inclusion
$B'\seq B$ where $B'$ is another $\ztwo$-graded Banach space. Suppose
that $(\A_t)$ restricts to a strongly continuous group $(\A_t)$ on $B'$, and
that
$\Delta\in\M^B(\L(B))$ restricts to an entire superfunction
$\Delta\in\M^{B'}(\L(B'))$.
Suppose also that there exists a function $C(\cdot)$ which is bounded on finite
intervals such that
\Beqn SelfSmEst
  \br\|\tilde\Delta(b)\|_{B'} \le C(\br\|b\|_{B})(1+\br\|b\|_{B'})
\Eeq
holds for all $b\in B'\seven$, and that the problem \eqref{AbstrIEq} is
$B$-complete. Then it is $B'$-complete, too. 
\end{lem}

\begin{proof}
Ad (i). For any $t_0\in I$ we have
\Beqn UndEqRelT0
 \phi(t) = \A_{t-t_0}\phi(t_0) + \int_{t_0}^t ds \A_{t-s}\tilde\Delta[\phi(s)].
\Eeq
Choose $K>0$ such that
$q(\A_t(b)) \le Kq(b)$ for all $b\in B\seven$, $t\in I$.
Setting
\Beq
 K':=\sup C(p(\tilde\Delta(\phi(t)))),\quad
 t_0:=\max\{t_2-1/(2KK'),\ (t_1+t_2)/2\},
\Eeq
\eqref{UndEqRelT0} implies for $t\in\left[t_0,t_2\right)$
\Beq
 q(\phi(t)) \le Kq(\phi(t_0)) + 1/2\cdot \bigl(1+ \max_{s\in\br[t_0,t]}
q(\phi(s))\bigr)
\Eeq
and hence
$1/2\cdot \max_{s\in\br[t_0,t]} q(\phi(s)) \le Kq(\phi(t_0))+1/2$,
showing that $\sup_{s\in\left[t_0,t_2\right)} q(\phi(s)) <\infty$.
The lower interval boundary is done analogously.

Ad (ii). This is an obvious corollary.
\end{proof}

\begin{proof}[Proof of Thm. \ref{IsSubSmf}]
Looking at the linear term of the Taylor expansion of $\Xi\sol$
at the origin we get that $\A_\mutarg$ maps continuously $E\Cau\to E$.
Now it is easy to check that the smf morphism $\alpha: \L(E)\to\L(E)$ given by
\Beq
 \alpha[\Xi]:=\Xi+ \Xi\sol[\Xi(0)]-\A_\mutarg\Xi(0)
\Eeq
makes the diagram
\begin{eqnarray*}
 &\L(E\Cau) & \\
 \A_\mutarg\Xi\Cau&\swarrow\hskip1cm\searrow&\Xi\sol \\
 \L(E) & \too{\qquad\alpha\qquad} &\L(E)
\end{eqnarray*}
commutative.  Also, we have a decomposition $E = E\zero \oplus E\free$ with
\Beq
  E\free := \{\xi\in E:\quad \xi(t)=\A_t\xi(0)\},\quad
  E\zero := \{\xi\in E:\quad \xi(0)=0\},
\Eeq
(both terms are equipped with the subspace topology)
with the corresponding continuous projections given by
$\pr\free(\xi):=\A_\mutarg\xi(0)$,\ \ $\pr\zero:=1-\pr\free$.
Therefore, the assertion follows once we have shown that $\alpha$ is an
automorphism.

We get an identification $\L(E) = \L(E\free) \times \L(E\zero)$,
with the corresponding projection morphisms being
$\L(\pr\free)$, $\L(\pr\zero)$, and $\alpha$ becomes the composite
\Beq
 \L(E) = \L(E\free) \times \L(E\zero)
  \too{(\Xi\sol\circ\pi)\times\L(\seq)}
  \L(E)\times \L(E) \too{\L(+)} \L(E)
\Eeq
where $\pi$ is the projection onto Cauchy data. As often in supergeometry,
it is convenient to look at the point functor picture, i.~e. we look how
$\alpha$ acts on $Z$-families of configurations: For any smf $Z$ we get a map
\Beqn Yoneda
 \opn Mor (Z,\L(E)) \to \opn Mor (Z,\L(E)),\quad \xi\mapsto\alpha\circ\xi,
\Eeq
and our assertion follows once we have shown that this is always
an isomorphism. (Indeed, it is sufficient to take $Z:=\L(E)$,\ \  $\xi:=\opn Id
$.)

Now $\alpha$ acts on $\xi\in\M^{E}(Z) = \opn Mor (Z,\L(E))$ by
\Beq
 \xi = \xi\free + \xi\zero\mapsto \Xi\sol[\xi\free(0)] + \xi\zero.
\Eeq
We show injectivity of \eqref{Yoneda}: If
$\alpha\circ\xi=\alpha\circ\xi'$
then, taking Cauchy data at both sides, we get that $\xi\free$,
$(\xi')\free$ have the same Cauchy data;
hence $\xi\free=(\xi')\free$, and the hypothesis now implies $\xi=\xi'$.

We show surjectivity of \eqref{Yoneda}: Given $\xi\in\M^{E}(Z)$, its
preimage is given by $\xi\free + \xi\zero$ with
\Beq
 \xi\zero:=\xi - \Xi\sol[\xi(0)],\qquad \xi\free:= \A_\mutarg \xi(0).
\Eeq
The Theorem is proved.
\end{proof}

The following is an abstract version of \cite[Thm. 3.4.3]{[CAUCHY]}.
The proof relies on \whatref{Prop. 2.4.2}.

\begin{thm}
Let be given a continuous, even inclusion
$E\seq B(\Bbb R)$ where $E$ is another $\ztwo$-lcs such that
the set of all linear forms on $E$ which arise by restricting elements of the
dual $B(\Bbb R)^*$ is strictly separating  (cf. \whatref{2.4}).

Let be given an smf $Z$ and a superfunction $\XiCauP\in\M^{E\Cau}(Z)$
(i.~e. a family of Cauchy data). Suppose that
\begin{itemize}
\item[(i)]
 for each $z\in Z$, there exists
 a solution $\phi_z(\cdot)\in E\seven(\mathbb R)$ of the underlying even
problem \eqref{UnderlEvProbl}
 with $\phi_z(0)=\XiCauP(z)$;
\item[(ii)]
 for each $z\in Z$, the power series $\Xi\sol_{\phi_z(0)}\in\P(B;B(\Bbb R))$,
 as defined by Lemma \ref{XiSolXl} and (i),
 restricts to a power series $\Xi\sol_{\phi_z(0)}\in\P(E\Cau;E)$.
\end{itemize}
Then there exists a unique $Z$-family of solutions $\Xi'\in\M^E(Z)$
which has $\XiCauP$ as its Cauchy data, i. e. $\Xi'(0) = \XiCauP$.
The Taylor expansion of $\Xi'$ at $z$ is given by
\Beqn DefXiZ
 \Xi'_z=\Xi\sol_{\phi_z(0)}[\XiCauP_z-\phi_z(0)]
\Eeq
where $\Xi\sol_{\phi_z(0)}$ is given by Thm. \ref{XiSolXl}. (Note that the
insertion is defined since the power series inserted has no absolute term.)

Also, the underlying map of the arising smf morphism
$\Xi':Z\to\L(E)$ is $z\mapsto\phi_z$.
\qed\end{thm}

One gets a general method for
showing solvability in function spaces:

\begin{cor}\label{GenSolvCrit}
Suppose that we are given a continuous, even inclusion
$E\seq B(\Bbb R)$ where $E$ is another $\ztwo$-lcs such that:
\begin{itemize}
\item[(i)]
 the problem \eqref{AbstrIEq} is $(B,E\Cau\seven)$-complete;
\item[(ii)]
 the set of all linear forms on $E$ which arise by restricting elements of the
dual
$B(\Bbb R)^*$ is strictly separating  (cf. \whatref{2.4});
\item[(iii)]
for $\phi\in E\Cau\seven$, the power series $\Xi\sol_{\phi}\in\P(B;B(\Bbb R))$,
as defined by Lemma \ref{XiSolXl} and (i), restricts to a power series
$\Xi\sol_{\phi}\in\P(E\Cau;E)$.
\end{itemize}

Then the problem \eqref{AbstrIEq} is solvable in $\L(E)$.
\qed\end{cor}

\subsection{Smoothness and support scales: the proofs}
\strut
\begin{proof}[Proof of Prop. \ref{SelfTmpSm}]
It is sufficient to show that for $j=0,\dots,l-1$,
$\Xi'\in \M^{C^j(I,B_{l-j})}(Z)$ implies
$\Xi'\in \M^{C^{j+1}(I,B_{l-j-1})}(Z)$.

By the Closed Graph Theorem, the generator $K$ is defined as a bounded
operator $K: B_{l-j} \to B_{l-j-1}$ for all $j$. Differentiation of
\eqref{AbstrIEq}
yields that \eqref{AbstrDEq} holds within $\M^{B_{l-1}(I)}(Z)$.

Clearly, we have $\Delta[\Xi']\in \M^{C^j(I,B_{l-j})}(Z)$ and
$K\Xi'\in \M^{C^j(I,B_{l-j-1})}(Z)$. Hence, the r.~h.~s. of \eqref{AbstrDEq}
lies in
$\M^{C^j (I,B_{l-j-1})}(Z)$, from which the assertion follows.
\end{proof}

\begin{proof}[Proof of Thm. \ref{CausUniqThm}]
W. l. o. g., we may assume $I=[0,t_0]$ with some $t_0>0$.
Also, we may suppose $Z$ to be a superdomain $Z\seq\L(F)$.
Supposing that our assertion is wrong, we can pick a $z\in Z$ such that
the set $\{t\in[0,t_0]:\ (\Xi'-\Xi")_z(s)\equiv_s 0 \ \text{for}\ s\in[0,t]\}$
is smaller than $I$. This set is easily seen
to be closed; let $t_2$ be its maximum.
{}From the hypotheses we get with $\Theta:=\Xi"-\Xi'$ that
\Beqn Theta1
 \Theta_z(t) \equiv_t
 \int_{t_2}^t ds \A_{t-s}(\Delta[\Xi'_z+ \Theta_z]- \Delta[\Xi'_z])(s)
\Eeq
for $t\in I$. Using the operator family $(\E_t)$ from the definition of support
scales, we get an even continuous linear
operator $\E: B([t_2,t))\to B([t_2,t))$, \ \ $(\E\xi)(t):=\E_t\xi(t)$.
Now \eqref{Theta1} yields
\Beqn Theta=F
 \Theta_z(t) \equiv_t
 \int_{t_2}^t ds \A_{t-s}(\Delta[\Xi'_z+ \E\Theta_z] - \Delta[\Xi'_z])(s).
\Eeq
Choose some $r\in\CS(F)$ such that
the relevant Taylor expansions $\Xi'_z,\Xi"_z$ lie in the Banach space
$\P(F,r;B(I))$. For shortness, we will write $\br\|\cdot\|_G$ for the norms in
$\P(F,r;G)$ where $G$ is one of the Banach spaces $B$,  $B([t_2,t])$, etc.
Using \eqref{EstIntAg} we get that with some $C_1>0$
\Beq
 \br\|\int_{t_2}^t ds \A_{t-s}(\Delta[\Xi'_z+ \E\Theta_z] -
\Delta[\Xi'_z])(s)\|_B
 \le C_1 \br|t-t_2|
 \br\| \Delta[\Xi'_z+ \E\Theta_z] - \Delta[\Xi'_z] \|_{B([t_2,t])}
\Eeq
for $t\in [t_2,t_0]$. Because of \eqref{Theta=F}s
and the estimate required for $\E_t$, this implies with some $C_2>0$
\Beqn ContOfE
 \br\|\E_t\Theta_z(t)\|_{B}
 \le C_2 \br|t-t_2|
 \br\| \Delta[\Xi'_z+ \E\Theta_z] - \Delta[\Xi'_z] \|_{B([t_2,t])}.
\Eeq
Let $\phi:=\Xi'_z(t_2)[0]\in B\seven$, and choose some $c>0$ such that
$\Delta_\phi\in\P(B,c\br\|\cdot\|;B)$.
(We deliberately do not make use of entireness of $\Delta$, which entails the
validity of this for any $c>0$).
Changing the norm in $B$, we may for notational convenience assume $c=1$.
Now choose $\epsilon>0$ such that for $t\in I':=\br[t_2,t_2+\epsilon]$ we have
\Beqn CausUnPfAbsT
 \br\|\Xi'_z(t)[0] - \phi\|_B < \frac14,\qquad \br\|\E\Theta_z(t)[0]\|_B <
\frac14
\Eeq
(this is possible since $\Xi'_z(t)[0],\E\Theta_z(t)[0]$  depend continuously on
$t$). Now, by dilating $r$, we may
assume
\Beqn CausUnPfRst
 \br\|\Xi'_z - \Xi'_z[0]\|_{B(I')} <\frac14,\qquad
 \br\|\E\Theta_z-\E\Theta_z[0]\|_{B(I')} <\frac14
\Eeq
(this is possible since both power series do not have an absolute term).
Now, letting $\Xi$ and $\delta\Xi$ be independent functional variables, we may
expand into bihomogeneous
components:
\Beq
 \Delta[\Xi+\phi+\delta\Xi]-\Delta[\Xi+\phi] = \sum_{i,j\ge0}
D_{(i,j)}[\Xi,\delta\Xi] \in \P(B\oplus B,\br\|\cdot\|;B),
\Eeq
with $D_{(i,0)}=0$ for all $i$. For arbitrary
$\Xi",\delta\Xi"\in\P(F,r;B)\seven$ we
get the estimate (cf. \CMPref{Proof of Prop. 3.3})
\Bmln CausUnPfEst
 \br\|\Delta[\Xi"+ \phi+\delta\Xi"]-\Delta[\Xi"+\phi]\|_B
 \\
 \le \sum_{i\ge0,\ j\ge1}  \br\|D_{(i,j)}[\Xi,\delta\Xi]\|_{\P(B\oplus
B,\br\|\cdot\|;B)}
  \br\|\Xi"\|^i_{B}
  \br\|\delta\Xi"\|^j_{B}
  \le \frac{C_3 \br\|\delta\Xi"\|_B}{(1- \br\|\Xi"\|_B)(1-\br\|\delta\Xi"\|_B)}
\Eml
with $C_3:= \br\|\Delta[\Xi+\phi+ \delta\Xi]-\Delta[\Xi+\phi]\|_{\P(B\oplus
B,\br\|\cdot\|;B)}$. Taking here
$\Xi":=\Xi'_z(t) - \phi$,
$\delta\Xi":= \E\Theta_z(t)$ with $t\in I'$, we get because of
\eqref{CausUnPfAbsT}, \eqref{CausUnPfRst} that
$\br\|\Xi"\|, \br\|\delta\Xi"\|< \frac12$, and \eqref{CausUnPfEst} yields
\Beqn EstDiffDelta
 \br\| \Delta[\Xi'_z(t)+ \E\Theta_z(t)] - \Delta[\Xi'_z(t)]\|_B   \le
 4 C_3 \br\|\E\Theta_z(t)\|_B
\Eeq
for $t\in I'$. Putting \eqref{ContOfE}, \eqref{EstDiffDelta}
together we get
\Beq
 \br\|\E\Theta_z(t)\|_{B} \le 4C_2C_3 \br|t-t_2|
 \br\|\E\Theta_z|_{[t_2,t]}\|_{B([t_2,t])}
\Eeq
for $t\in I'$. Now, for (say) $0<\br|t-t_2| < 1/(8C_2C_3)$, this estimate
implies
$\br\|\E\Theta_z(t)|\|_B=0$, in contradiction to the choice of $t_2$.
\end{proof}

\begin{proof}[Proof of Lemma \ref{DeltaLem}]
Uniqueness is easy to see. We construct
$\Delta \in \M^{C^l(M,B)}\alb (\L\alb(C^l(M,B)))$
by specifying its Taylor expansions:
\Beqn DeltaPhiAss
 C^l(M,B\seven)\owns\phi \mapsto \Delta_\phi\in\P(C^l(M,B); C^l(M,B)),
\Eeq
\Beqn DeltaPhiT
 (\Delta_\phi)_{(r|s)}(\phi_1,\dots,\phi_r|\psi_1,\dots,\psi_s)(t) :=
 (\Delta_{\phi(t)})_{(r|s)}(\phi_1(t),\dots,\phi_r(t)|
  \psi_1(t),\dots,\psi_s(t))(t)
\Eeq
for $\phi_i\in C^l(M,B\seven),\ \  \psi_i\in C^l(M,B\sodd),\ \ t\in M$.

For showing well-definedness, we remark that the topology of
$C^l(M,B)$ is defined by seminorms of the form
$\br\|\phi\|_{K,l}:=\sum_{|\nu|\le l} \sup_{t\in K}
\br\|\partial^\nu\phi(t)\|/\nu!$
where $K\Subset M$ is contained in a coordinate patch identifying it
with the unit ball of $\Bbb R^n$ (the modification for $l=\infty$ is obvious).

For $b\in B$ and $K$ as above, choose $U^b\in\CB(B)$  such that
$\Delta_b\in \P(B,4U^b; C^l(K,B))$. Now, given $K$ and $\phi\in C^l(M,B)$,
compactness
allows to find $t_1,\dots,t_N\in M$ such that
$\phi(K) \seq \bigcup_{i=1}^N (\phi(t_i) + U^{\phi(t_i)})$; set
$U^{K,\phi} := \bigcap_{i=1}^N U^{\phi(t_i)}$.

Now, given $t\in K$, choose $i$ with $\phi(t)-\phi(t_i)\in U^{\phi(t_i)}$. It
follows that
\Beq
 \Delta_{\phi(t)} =\opn t _{\phi(t)-\phi(t_i)} \Delta_{\phi(t_i)} \in
 \P(B,2U^{\phi(t_i)}; C^l(K,B)) \seq \P(B,2U^{K,\phi}; C^l(K,B)).
\Eeq
We get a map
\Beqn TheClMap
 K \owns t\mapsto \Delta_{\phi(t)} \in\P(B,2U^{K,\phi}; C^l(K,B)).
\Eeq
Moreover, for each $i$, the map
\Beq
 \phi(t_i) + U^{\phi(t_i)} \to  \P(B,2U^{\phi(t_i)}; C^l(K,B)),
 \quad \phi(t_i) +b\mapsto \Delta_{\phi(t_i)+b}=\opn t _b\Delta_{\phi(t_i)}
\Eeq
is real-analytic. It follows that the composite map \eqref{TheClMap} is $C^l$.

Hence, given $r,s,\phi_1,\dots,\phi_r,\psi_1,\dots,\psi_s$, the r.~h.~s. of
\eqref{DeltaPhiT} depends in a $C^l$ way on $t\in K$; since this is true
for all $K$, it follows that $\Delta_\phi$ is well-defined as a formal power
series.

Now, for $\br|\nu|\le l$ and
$\phi_i\in C^l(M,U^{K,\phi}\cap B\seven),\ \  \psi_i\in C^l(M,U^{K,\phi}\cap
B\sodd),\ \ t\in K$,
\Beq
 \frac1{\nu!} \br\| \partial^\nu
(\Delta_{\phi(t)})_{(r|s)}(\phi_1(t),\dots,\phi_r(t)|
  \psi_1(t),\dots,\psi_s(t))(t)\|
\le
 2^{-r-s}\br\| \Delta_{\phi(\cdot)}\|_{K,l},
\Eeq
where $\br\| \Delta_{\phi(\cdot)}\|_{K,l}$ is the $C^l$ norm of the map
\eqref{TheClMap}.
Hence
\Beq
\br\| \Delta_{\phi} \|_{\P(C^l(M,B), C^l(M,U^{K,\phi}); C^l(K,B))}
\le
 \sum_{i=0}^l (\opn dim M)^i \cdot \sum_{r,s} 2^{-r-s} \br\|
\Delta_{\phi(\cdot)}\|_{K,l} <\infty
\Eeq
which proves that $\Delta_\phi$ is an analytic power
series, i.e. \eqref{DeltaPhiAss} is well-defined.
Now one applies the strictly separating family  (cf. \whatref{2.4}) of linear
functionals
$C^l(M,B)\to \Bbb R$, $\phi\mapsto \br<b^*,\phi(t)>$ where
$b^*\in B^*$ and $t\in M$ to conclude that
the map \eqref{DeltaPhiAss} is a superfunction.
\end{proof}

\subsection{Application: the proofs}
\begin{proof}[Proof of Prop. \ref{SobSlfSm}]
Of course, we can assume $l=1$. Fix $a\in\{1,\dots,d\}$. Using the algebra
property of the Sobolev
spaces (cf. \cite{[Hd2IsAlg]}), there is a constant $K_1$ such that
\Beq
  \br\|\partial_a\tilde\Delta_i[\phi]\|_{H_{k+\sd_i}(\rdm)}
  \le K_1
  \sum_{j,\nu}
  \br\|\partial_a\partial^\nu\phi_j\|_{H_{k+\sd_i}(\rdm)}
  \cdot
 \br\|\frac\partial{\partial(\partial^\nu\ul\Xi_j)} \tilde\Delta_i[\phi]\|
 _{H_{k+\sd_i}(\rdm)}
\Eeq
where, because of \eqref{SmCond}, the sum runs over those $j=1,\dots,N\seven$
and $\nu\in {\Bbb Z_+}^d$
for which $\sd_i\le \sd_j-\br|\nu|$. This restriction implies
\Beq
 \br\|\partial_a\partial^\nu\phi_j\|_{H_{k+\sd_i}(\rdm)}
 \le
 \br\|\partial_a\phi_j\|_{H_{k+\sd_j}(\rdm)} \le
 \br\|\phi\|_{\cH\V_{k+1}}.
\Eeq
Setting $B:=\cH\V_k$, $B':=\cH\V_{k+1}$, and
\Beq
 C(r):=K_2 \cdot \sup_{\phi\in\cH\V_k,\ \ \br\|\phi\|\le r}
  \br\|\tilde\Delta[\phi]\|_{\cH\V_k},
\Eeq
with suitable $K_2>0$, \eqref{SelfSmEst} is satisfied, and the assertion
follows from Lemma \ref{SelfSm}.(ii).
\end{proof}

\begin{proof}[Proof of Thm. \ref{CausCauUniqu}]
We first note:

\begin{lem}
Let be given a $Z$-family $\Xi'\in\M^{\cH\V_k(I)}(Z)$
with $k>d/2$, and suppose that
\Beq
 L_i[\Xi']|_{\Omega(p)} =0 \quad (i=1,\dots, N)
\Eeq
within $\M^{\cD'(\Omega(p))\otimes\Bbb R^{N\seven|N\sodd}}(Z)$.
Then $\Xi'$ satisfies the integral equation
\Beq
 \Xi'(t,y) = \A_t\Xi'(0)(y) + \int_0^t ds \A_{t-s}\Delta[\Xi'](y)
\Eeq
within $\O^{\Bbb R^{N\seven|N\sodd}}(Z)$ for all $(t,y)\in\Omega(p)$.
\qed\end{lem}

Let $p=(s,x)\in\rdmm$, and assume $s>0$ ($s<0$ is done mutatis mutandis):
Within $\cH\V_k$, we have a support scale $(\spsc_t)_{t\in [0,s]}$,
\Beq
 \spsc_t :=\left\{\xi\in\cH\V_k:\quad \supp \xi \cap
\J((x,s-t))=\emptyset\right\}.
\Eeq
The Theorem now follows from Thm. \ref{CausUniqThm}.
\end{proof}


Before proceeding, we do some technical preparations. We will use the
notations
\Beq
 \cEV := C^\infty(\rdmm,\Bbb R^{N\seven|N\sodd}),\quad
 \cECauV := C^\infty(\rdm,\Bbb R^{N\seven|N\sodd}),
\Eeq
\Beq
 \cEcV :=C^\infty_c(\rdmm,\Bbb R^{N\seven|N\sodd}),\quad
 \cEcCauV:= C^\infty_c(\rdm,\Bbb R^{N\seven|N\sodd}).
\Eeq
We need a technical notion: Given a seminorm
$p\in\CS(\cD(\rdmm))$, we define the {\em support of $p$}, denoted by $\supp
p$,
as the complement of the set of all $x$ which have a neighbourhood
$U\owns x$ such that $\supp\varphi\seq U$ implies $p(\varphi)=0$.
Obviously, $\supp p$ is closed; using partitions of unity one shows that
$\supp\varphi\seq \rdmm\setminus\supp p$ implies $p(\varphi)=0$.

For every $p\in\CS(C^\infty(\rdmm))$, $\supp p$ is compact (where we
have silently restricted $p$ to $\cD(\rdmm)$). On the other hand,
given $p\in\CS(\cEcV)$, the set $\supp p\cap\bV_r$ (cf. \eqref{bVr})
is compact for all $r\ge0$.

Given a bounded open set $\Omega\Subset\rdmm$, we denote by
$\J(\Omega)\Subset\rdm$ the {\em causal influence domain of $\Omega$}
on the Cauchy hyperplane, i.~e. the set of all $x\in\rdm$ such that
$(0,x)$ lies in the twosided light cone of a point in $\Omega$.

For $\Omega\Subset\rdmm$, $l\ge0$, define the seminorm
$q_{l,\Omega}\in\CS(\cEV)$ by
\Beq
 q_{l,\Omega}(\xi) = \sum_{i=1}^N
 \sup_{(t,x)\in\Omega} \sum_{\nu\in\Bbb Z_+^{d+1},\ |\nu|\le l}
 \br|\partial^\nu \xi_i(t,x)|;
\Eeq
thus $\supp q = \Omega$. Also, for $J\Subset\rdm$, $k\ge0$, define
the seminorm $p_{k,J}\in\CS(\cECauV)$ by
\Beq
 p_{k,J}(\xi\Cau) := \sum_{i=1}^N \sup_{x\in J}
 \sum_{\nu\in\Bbb Z_+^d,\ |\nu|\le k} \br|\partial^\nu \xi\Cau_i(x)|;
\Eeq
thus, $\supp p_{k,J}=J$.

\begin{proof}[Proof of Thm. \ref{cDMainThm}]
\begin{lem}\label{CausInfEst}
Under the hypotheses of Thm. \ref{cDMainThm},
fix a Cauchy datum $\phi\in\cEcCauVs$.

(i) There exists a unique element $\phi'\in\cEcVs$ with
$\phi'(0)=\phi$ which solves the underlying system \eqref{UnderlSyst}.

(ii) For $\Omega\Subset\rdmm$, $l\ge0$, let
$k>\smloss l+d/2 + \max\{\sd_1,\dots,\sd_N\}$.
Then, for all $\epsilon>0$,
the power series $\XiCData[\Xi\Cau]$
given by Lemma \ref{XiSolXl} satisfies a
$(q_{l,\Omega}, C_\epsilon p_{k,J_\epsilon})$-estimate
(cf. \CMPref{3.1}) with some $C_\epsilon>0$, where
$J_\epsilon=U_\epsilon(\J(\Omega))$ is the
$\epsilon$-neighbourhood of $\J(\Omega)$.

(iii) Let $q\in\CS(\cEV)$ be arbitrary. Then there exists $k>0$ such that
for all $\epsilon>0$, $\XiCData[\Xi\Cau]$ satisfies the
$(q,C_\epsilon p_{k,J_\epsilon})$-estimate with some $C_\epsilon>0$,
where $J_\epsilon=U_\epsilon(\J(\supp q))$.

(iv)
$\XiCData[\Xi\Cau]$ is an analytic power series from
$\cEcCauV$ to $\cEcV$:
\Beq
 \XiCData[\Xi\Cau]\in\P(\cEcCauV; \cEcV)\sevR.
\Eeq
\end{lem}

\begin{proof}
Ad (i).
{}From the completeness hypothesis and Cor. \ref{ComplSystCor}, we get
a solution $\phi'\in(\cH\V_k)\seven$ of \eqref{UnderlSyst} with
$\phi'(0)=\phi$. Now Prop. \ref{SobSlfSm}
and Thm. \ref{CausCauUniqu}.(ii) together with the Sobolev Embedding Theorem
yield $\phi'\in\cEcVs$.

Ad (ii).
Let $I\seq\Bbb R$ be the projection of $\Omega$ onto the time axis.
By the Sobolev Embedding Theorem, there exists a constant $C_1$ such that
\Beq
 q_{l,\Omega}(\varphi) \le  C_1 \cdot\br\|\varphi|_I\|_{\cH_k^{\opn V ,l}(I)}
\Eeq
for $\phi\in\cH_k^{\opn V ,l}(I)$ where $\cH_k^{\opn V ,l}(I)$ is $B^l(I)$ with
$B:=\cH\V_k$
(cf. \eqref{DefBUpJ}).
Combining this with the Sobolev analyticity of
$\XiCData[\Xi\Cau]$ given by Lemma \ref{XiSolXl}, there exists a
constant $C_2$ such that we have for $r,s\ge0$,
$\varphi^1,\dots,\varphi^r\in\cEcCauVs$,
$\psi^1,\dots,\psi^s\in\cEcCauVso$
\Beq
 q_{l,\Omega}\Bigl(\Bigl<\br(\XiCData)_{r|s},
  \bigotimes_{m=1}^r \varphi^m\otimes \bigotimes_{n=1}^s \Pi \psi^n\Bigl>\Bigr)
 \le
 C_2\cdot\prod_{m=1}^r \br\|\varphi^m\|_{\cH\V_k}\cdot
  \prod_{n=1}^s\br\|\psi^n\|_{\cH\V_k}
\Eeq
(cf. \CMPref{3.1} for the notation on the l.~h.~s.).
Now choose some buffer function $h\in\cD'(\rdm)$ with $\supp h\seq J_\epsilon$
and $h|_{\J(\Omega)}=1$. By causality (cf. Thm. \ref{CausUniqThm}.(ii)),
we have $\XiCData[\Xi\Cau]|_\Omega = \XiCData[h\Xi\Cau]|_\Omega$, and hence
\Bal
 q_{l,\Omega}\Bigl(\Bigl<\br(\XiCData)_{r|s},
 \bigotimes_{m=1}^r \varphi^m\otimes
 \bigotimes_{n=1}^s \Pi \psi^n\Bigr>\Bigr)
 &=
 q_{l,\Omega}\Bigl(\Bigl<\br(\XiCData)_{r|s},
 \bigotimes_{m=1}^r (h\varphi^m)\otimes
 \bigotimes_{n=1}^s \Pi(h\psi^n)\Bigr>\Bigr)\\
 &\le
 C_2\cdot\prod_{m=1}^r
 \br\|h\varphi^m\|_{\cH\V_k}\cdot \prod_{n=1}^s\br\|h\psi^n\|_{\cH\V_k}
\Eal
But obviously $\br\|h\cdot\|_{\cH\V_k}$ is estimated from above by
$C_\epsilon p_{k,J_\epsilon}(\cdot)$ with some $C_\epsilon>0$,
and the assertion follows.

Ad (iii).
Since the collection of all $q_{l,\Omega}$ defines the topology
of $\cEV$, there exist $l,C'$,and $\Omega'\Subset\rdmm$ such that
$q\le C'q_{l,\Omega'}$.
However, $\Omega'$ may be larger than $\supp q$. Choose a buffer
function $g\in\cD(\rdmm)$,\ \  $g\ge0$, with $g|_{\supp q}=1$,\ \
$\supp g\seq J_{\epsilon/2}$. Then
\Beq
 q(\cdot) = q(g\cdot) \le C'q_{l,\Omega'}(g\cdot)
 \le C'_\epsilon q_{l,J_{\epsilon/2}}(\cdot)
\Eeq
with some $C'_\epsilon>0$. The assertion now follows from (ii).

Ad (iv).
Let be given a seminorm $q\in\CS(\cEcV)$. With standard methods one
constructs for $i>0$ buffer functions $f_i\in C^\infty(\rdmm)$
with $f_i|_{\bV_{i-1}}=0$, $f_i|_{\rdmm\setminus\bV_i}=1$
(cf. \eqref{bVr}). Set for convenience $f_0:=1$.
For the seminorms $q_i:=q((f_i - f_{i+1})\cdot)\in\CS(\cEcV)$ we get
\Beqn pIsSumQ
 q(\varphi) \le \sum_{i\ge0} q_i(\varphi)
\Eeq
for all $\phi\in\cEcV$, where in fact only finitely many terms
on the r.~h.~s. are non-zero. Now
\Beq
 \supp q_i\seq \bV_{i+1} \cap \supp q
\Eeq
which is 
compact. Also, for $i\ge1$, we have
$(f_i - f_{i+1})|_{\bV_{i-1}}=0$ and hence
\Beqn OutOfVi-1
 \supp q_i \cap \bV_{i-1}=\emptyset.
\Eeq
Because of \eqref{OutOfVi-1}, we have
$\J(\supp q_i)\seq \{x\in\rdm: \ \ \br\|x\|\ge i-1\}$ for $i\ge1$; hence,
setting $J_i:= \{x\in\rdm: \ \ \br\|x\|\ge i-2\}$,
Lemma \ref{CausInfEst}.(ii) yields for each $i$ numbers
$C_i>0$,\ \ $k_i\ge0$ such that $\XiCData[\Xi\Cau]$ satisfies a
$\br( q_i, C_i p_{k_i,J_i})$-estimate.

It follows that for each $\varphi\in\cEcCauV$, the sum
\Beq
   p(\varphi) := \sum_i C_i p_{k_i,J_i}(\varphi)
\Eeq
has only finitely many nonvanishing terms; using
\cite[Thm. 15.4.1]{[Hormander]}, we have $p:=p(\cdot)\in\CS(\cEcCauV)$.
It follows directly from the definition of the $(q,p)$-estimates
(cf. \CMPref{3.1}) and \eqref{pIsSumQ} that the
$\br( q_i, C_i p_{k_i,J_i})$-estimates
for $\XiCData[\Xi\Cau]$ imply the $(q,p)$-estimate wanted.

The Lemma is proved.
\end{proof}

Thm. \ref{cDMainThm} now follows from Cor. \ref{GenSolvCrit}.
\end{proof}

\begin{proof}[Proof of Thm. \ref{MainThmSm}]

\begin{lem}
Suppose that the problem \eqref{TheSyst} is causal.
Given a bosonic Cauchy datum $\phi\in\cECauVs$,
there exists a solution power series
$\XiCData[\Xi\Cau]\in\P(\cECauV;\cEV)\sevR$ such that
\Beqn CDtaWR
  \XiCData[\Xi\Cau](0) = \Xi\Cau+\phi.
\Eeq
\end{lem}

\begin{proof}
Choose a sequence of compactly supported bosonic Cauchy data
$\phi_{(n)}\in\cEcCauVs$,\quad $n\in\Bbb Z_+$,
such that $\phi_{(n)}|_{\ball n^d} = \phi|_{\ball n^d}$
for all $i$. Composing the power series
$\Xi\sol_{\phi_{(n)}}\in\P(\cEcCauV;\cEcV)$
given by Lemma \ref{CausInfEst}
with the projection
$\cEcV\to C^\infty(\ball n^{d+1})\otimes\Bbb R^{N\seven|N\sodd}$
we get a sequence of power series
\Beq
 \Xi_{(n)} := \Xi\sol_{\phi_{(n)}}|_{\ball n^{d+1}}
 \in\P(\cEcCauV; C^\infty(\ball n^{d+1})\otimes \Bbb R^{N\seven|N\sodd})\sevR.
\Eeq
Because of Thm. \ref{CausCauUniqu}.(i), the restrictions of
$\Xi_{(n+1)}$ and $\Xi_{(n)}$ onto $\ball n^{d+1}$ coincide. Hence there
exists a power series $\XiCData[\Xi\Cau]\in\P(\cEcCauV;\cEV)$ whose
restriction onto $\ball n^{d+1}$ is $\Xi_{(n)}$.
It is clear that this is a solution power
series which satisfies \eqref{CDtaWR}; the fact that it is actually
analytic with respect to the source space $\cECauV$ follows from
Lemma \ref{CausInfEst}.(iii).
\end{proof}

Now one proves quite analogously to the compactly supported case that the power
series
$\XiCData[\Xi\Cau]$ fit together to the superfunction
$\Xi\sol\in\M^{\Ci}(\L(\CiCau))$ wanted,
as well as the remaining assertions.
\end{proof}

\vfill

{\sc Technische Universit\"at Berlin}

Fachbereich Mathematik, MA 7 -- 2

Stra\ss e des 17. Juni 136

10623 Berlin

FR Germany

\medskip

{\em E-Mail  address: } schmitt@math.tu-berlin.de
\end{document}